\documentclass[journal,a4paper]{IEEEtran}
\usepackage{amsmath,amsfonts}
\usepackage{algorithmic}
\usepackage{algorithm}
\usepackage{array}
\usepackage{textcomp}
\usepackage{stfloats}
\usepackage{url}
\usepackage{cite}
\usepackage{verbatim}
\usepackage[pdftex]{graphicx}
\usepackage{color}
\usepackage{transparent}
\usepackage{import}
\usepackage{subfigure}
\usepackage{balance}
\usepackage{multirow}
\usepackage{cleveref}
\DeclareGraphicsExtensions{.eps}
\usepackage{epstopdf}
\usepackage{epsfig}
\usepackage{gensymb}
\usepackage{float} 
\UseRawInputEncoding
\def\BibTeX{{\rm B\kern-.05em{\sc i\kern-.025em b}\kern-.08em
    T\kern-.1667em\lower.7ex\hbox{E}\kern-.125emX}}
\usepackage{balance}
\usepackage{amsmath,amsfonts,amssymb}
\begin{document}

\title{Impacts of Real Hands on 5G Millimeter-Wave Cellphone Antennas: \\ Measurements and Electromagnetic Models}

\author{Bing Xue, Pasi Koivum\"aki, Lauri V\"ah\"a-Savo, Katsuyuki Haneda, ~\IEEEmembership{Member,~IEEE}, Clemens Icheln, 
\thanks{Manuscript received April XX, 2022; revised August XX, 2022.}
\thanks{B. Xue, P. Koivum\"aki, L. V\"ah\"a-Savo, K. Haneda and C. Icheln  are with the Department of Electronics and Nanoengineering, Aalto Universit$y$-School of Electrical Engineering, Espoo FI-00076, Finland (e-mail: bing.xue@aalto.fi).}}
\markboth{Journal of \LaTeX\ Class Files,~Vol.~14, No.~8, August~2021}%
{Shell \MakeLowercase{\textit{et al.}}: A Sample Article Using IEEEtran.cls for IEEE Journals}

\maketitle

\begin{abstract}
Penetration of cellphones into markets requires their robust operation in time-varying radio environments, especially for millimeter-wave communications. Hands and fingers of a human cause significant changes in the physical environments of cellphones, which influence the communication qualities to a large extent. In this paper, electromagnetic models of real hands and cellphone antennas are developed, and their efficacy is verified through measurements for the first time in the literature. Referential cellphone antenna arrays at $28$ and $39$~GHz are designed. Their radiation properties are evaluated through near-field scanning of the two prototypes, first in free space for calibration of the antenna measurement system and for building simplified models of the cellphone arrays. Next, radiation measurements are set up with real hands so that they are compared with electromagnetic simulations of the interaction between hands and simplified models of the arrays. The comparison showed a close agreement in terms of spherical coverage, indicating the efficacy of the hand and antenna array models along with the measurement approach. The repeatability of the measurements is $0.5$~dB difference in terms of cumulative distributions of the spherical coverage at the median level.

\end{abstract}

\section{Introduction}
\IEEEPARstart{D}{emands} of mobile users for high data-rate communications and access to information through wireless channels have always been increasing. Sub-6 GHz radio frequencies (RFs) cannot serve all possible scenarios of radio communications, especially the high data-rate communications for a high density of mobile users. Therefore, millimeter-wave communications attract the attention of the research community in recent years. Among the two operated bands of the frequency range of 5G cellular, the frequency range 2 (FR2) spanning between 24.5 to 43 GHz imposes more stringent challenges to ensuring robust operations in dynamic environments than the frequency range 1 (FR1) which covers lower frequency band \cite{3gpp2018}. One of the challenges is the sensitivity of antennas to the radio environment \cite{Helander2016, Hong2017, Xu2018, Krogerus2007}. When antennas and arrays are covered by dielectric objects such as bodies, fingers, and hand palms of the device operator, significant changes in their radiation properties are inevitable because of blockage and absorption of radiated fields. It is thus a goal of antenna designers of communication devices to minimize the detrimental effects of close-by objects on antennas \cite{Krogerus2007}. There are papers studying the antenna-human interaction at above-6 GHz RF, including FR2 \cite{Syrytsin2017,Syrytsin2017_2,Virk2020,Zhao2016,Hejselbaek2017,Zhao2020,Haneda2018,Haneda2018_2,Mikko2019,Lauri2021,Liu2021,Liu2022,Christian2021,Mikko2016, Raghavan2018, Raghavan2019, Zhekov2020, Nguyen2020, Raghavan2022,pasi2022,Yu2020}. These papers included works reporting numerical models of humans to run full-wave simulations of antenna-human interaction, e.g.,~\cite{Lauri2021, Christian2021}, and those measuring link blockage by human bodies, e.g.,~\cite{Virk2020, Haneda2018, Haneda2018_2}. They mainly cover human blockage effects at far fields of antennas, while papers~\cite{Mikko2016, Raghavan2018, Raghavan2019, Zhekov2020, Nguyen2020, Raghavan2022} discussed near-field effects of a human body on cellphone antenna radiations such as fingers’, hand palms’, or the whole hands' effects. For a repeatable study of hand effects, \cite{pasi2022} and \cite{Yu2020} established real hand models by laser scanning and photogrammetry that can be used in full-wave simulations of antenna-hand interaction. The overarching problem of the existing works reporting antenna-hand interaction is, however, {\it lack of comparisons between simulations and measurements} to cross-validate the simulation models and measurement methods. 
This paper continues the work \cite{pasi2022}, which reports three-dimensional (3D) models of real hands. We compare measured and simulated antenna-hand interaction based on the 3D models. The innovations reported in this paper are as follows.
\begin{enumerate}{}{}
\item Designs, fabrication, and measurements of referential dual-polarized linear antenna arrays on cellphone-sized chassis at 28 and 39 GHz, which serve for the antenna-hand interaction studies;
\item Setups of antenna-hand interaction measurements showing an acceptable degree of repeatability for a limited amount of measurement time, despite involving non-repeatable real human hands; and finally,
\item Cross-verified electromagnetic simulation models and measurement methods of the antenna-human interaction analysis at 28 and 39 GHz; the simulation models include 3D hand models of real humans while the measurements are with real human hands from which the 3D hand models are built. 
\end{enumerate}

The rest of the paper is arranged as follows: Section~\ref{sec:antenna} introduces our designs of referential dual-polarized antenna arrays for $28$ and $39$~GHz, including antenna elements, feed lines and cable connectors enclosed in a cellphone chassis. Section~\ref{sec:simulation} summarizes human hand models and their integration with the antenna array models. Section~\ref{sec:free-space} first introduces impedance matching and isolation characteristics of the manufactured antenna array. Then the principle of radiation measurements for antenna arrays is elaborated where an approach to de-embedding the losses of measurement setups is introduced. Section~\ref{sec:antenna-hand} details measurement setups with real hands, paying attention to the fact that each measurement must be completed while a subject human hand stays still. The measurements give evidence for the cross-validation of the simulation models and measurement approaches at the two frequencies. Finally, conclusions are summarized in Section~\ref{sec:conclusion}.

\section{Referential Millimeter-wave Cellphone Antenna Array}
\label{sec:antenna}
In this section, the designs and fabrications of two referential dual-polarized patch antenna arrays on a cellphone chassis are introduced. They are used for experimental antenna-hand interaction studies. 
\subsection{Antenna Element Design}
\begin{figure*}[htbp] 
    \centering
	\subfigure[]{
	\includegraphics[width=0.32\linewidth]{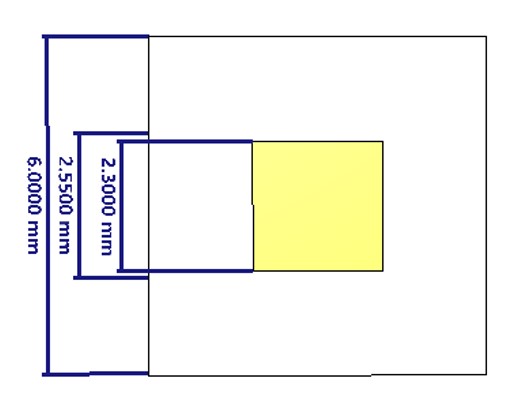}}
	\subfigure[]{
	\includegraphics[width=0.32\linewidth]{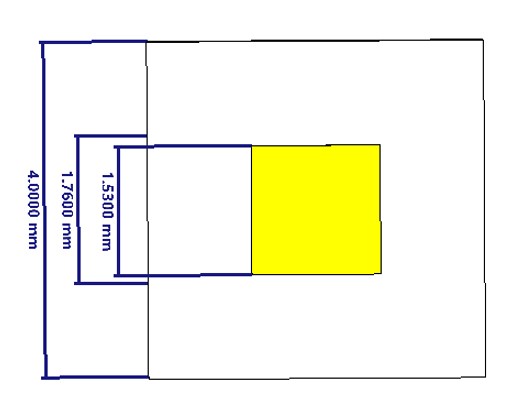}}
	\subfigure[]{
	\includegraphics[width=0.32\linewidth]{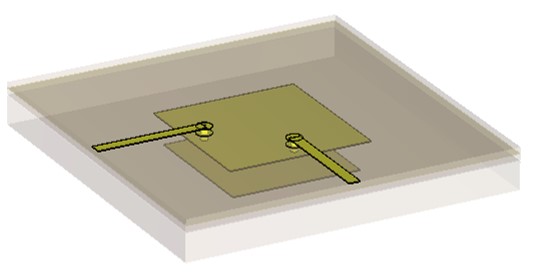}}
	\caption{(a) Dimensions of a stacked patch antenna element for 28 GHz, top view; (b) the same at 39 GHz. (c) Dual-polarization microstrip-feed structures, bottom view, with transparent ground plane and substrates.}
	\label{element}
\end{figure*}

Millimeter-wave cellphone antennas are usually designed using patches~\cite{Lauri2021, Mikko2016} due to their directivities illuminating the half-hemisphere and possibilities of dual-polarized arrays with suitable isolation between antenna ports. Two antenna arrays are designed to cover $28$ and $39$~GHz bands, as representatives of the FR2; a single antenna array cannot cover the two bands because patch antennas usually cover only up to $10$~\% relative bandwidth. Still stacked patches are used for the wide-band impedance matching at the two bands. Figs.~\ref{element} (a) and (b) show the antenna element's dimensions at 28 and 39 GHz while Fig.~\ref{element} (c) depicts the schematic of the antenna, seen from its bottom side implementing a ground plane. The dimension of the patch antenna along with the diameter of void rings for vias was optimized for the desired impedance bandwidth.

\subsection{Antenna Array Design and Fabrication}
\begin{figure*}[ht]
\centering
\includegraphics[width=0.6\linewidth]{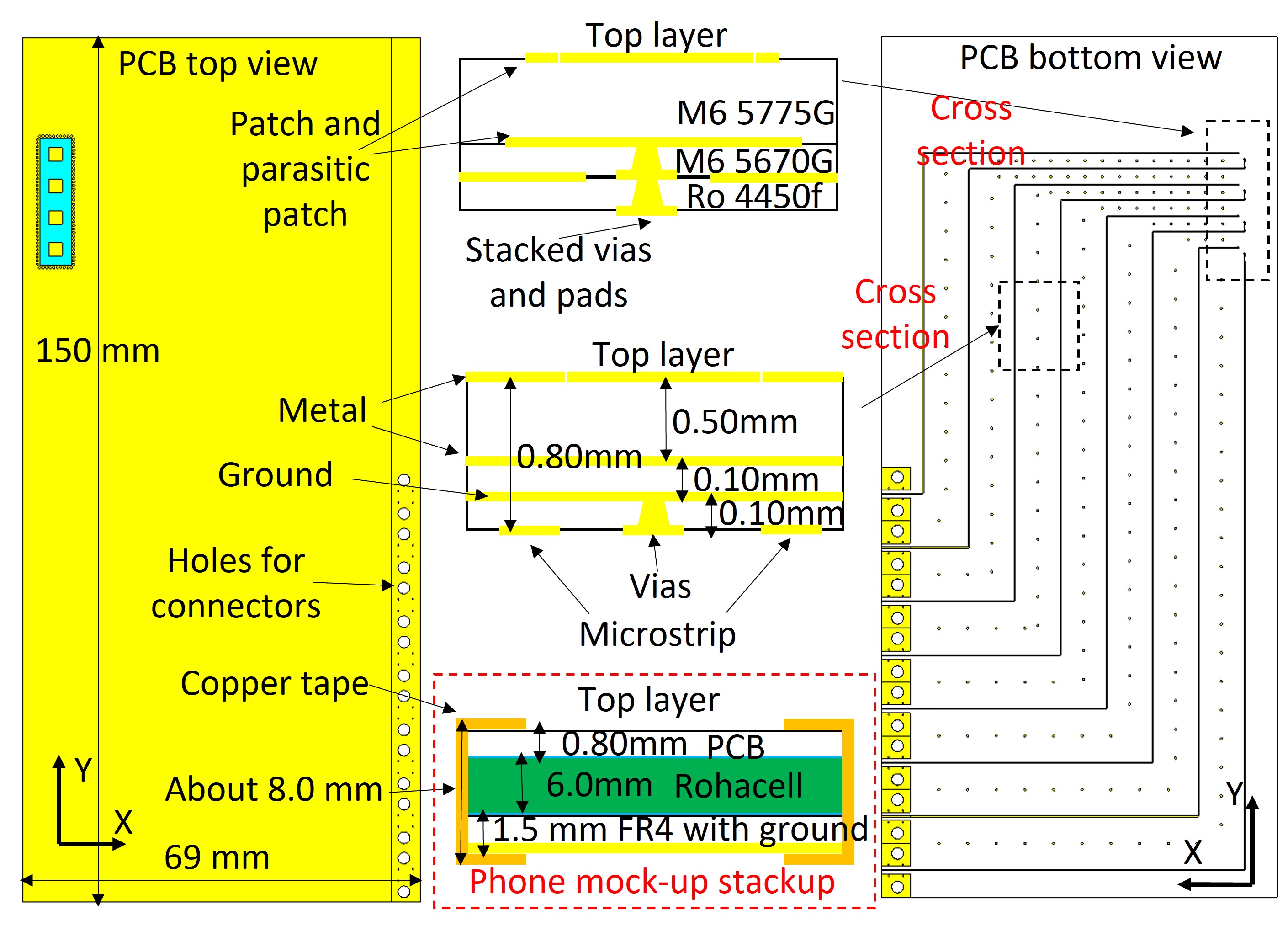}
\caption{Top and bottom views of the antennas and feeding lines at 28 GHz. Cyan and white parts represent the substrates while yellow parts are metal. A cellphone mock-up, consisting of the PCB, a rohacell and FR4 substrate, is shown in the stackup indicated by the red dashed rectangle.}
\label{28Phonearrange}
\end{figure*}

\begin{figure*}[ht]
\centering
\includegraphics[width=0.61\linewidth]{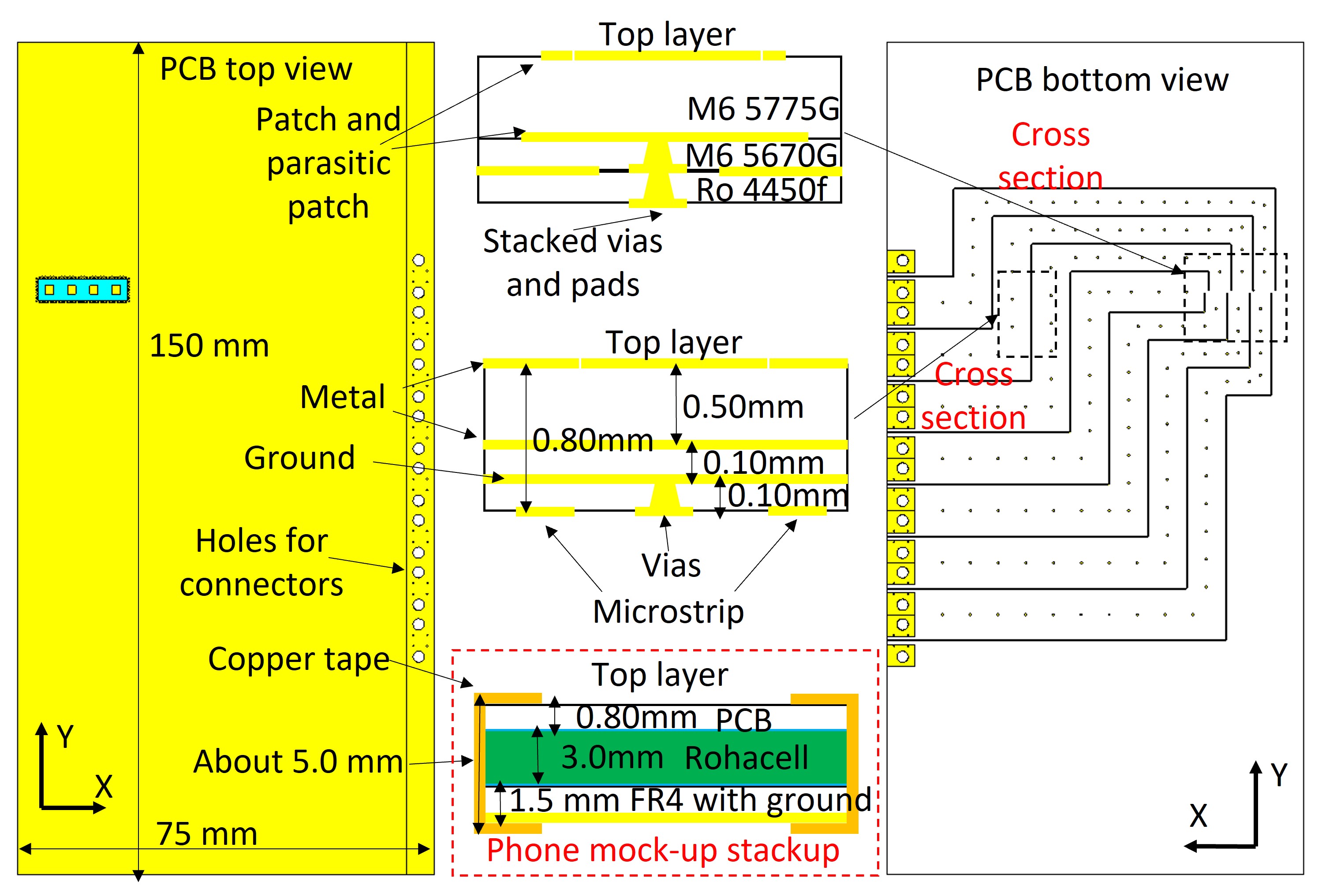}
\caption{ Top and bottom views of the antennas and feeding lines at 39 GHz. Cyan and white parts represent the substrates while yellow parts are metal. A cellphone mock-up, consisting of the PCB, a rohacell and FR4 substrate, is shown in the stackup indicated by the red dashed rectangle.}
\label{39Phonearrange}
\end{figure*}

The dual-polarized stacked patches are used to form $4$-element linear antenna arrays on a $150 \times 69~{\rm mm}^2$ printed circuit board (PCB) for 28 GHz and on a $150 \times 75~{\rm mm}^2$ PCB for 39 GHz as illustrated in Figs.~\ref{28Phonearrange} and \ref{39Phonearrange}, respectively. Each element has two ports, of which the main polarization is orthogonal to each other. The PCB consists of three layers of substrates, i.e., $0.5$ mm-thick top substrate (MEGRTON6 5775G: $\epsilon_{\rm r } = 3.62$, loss tangent: $0.005$ at 28 GHz and $0.006$ at 39 GHz), $0.1$ mm-thick middle substrate (MEGRTON6 5670G: $\epsilon_{\rm r} = 3.22$, loss tangent: $0.005$ at 28 GHz and $0.006$ at 39 GHz) and $0.1$ mm-thick bottom one (Rogers 4450f: $\epsilon_{\rm r} = 3.70$, loss tangent: $0.004$). The thickness of copper is $35$ $\mu$m for the top and bottom substrates while it is $18$ $\mu$m for the middle due to the manufacturer’s ability. The vias surrounding the antenna array connect the metal layers of the top and middle substrates to reduce surface waves on the PCB. Long feeding lines have to be used to connect the antennas to 8 end-launch connectors (\textit {Southwest Microwaves} 2.40 mm narrow-type) to minimize radiation from the connectors. The feed line width is $0.20$ mm while via pads and laser-drilled microvias are $0.40$ and $0.15$ mm in diameter, respectively, as shown in the cross-section schematic of Figs.~\ref{28Phonearrange} and \ref{39Phonearrange}. A void ring of $0.30$-mm diameter is used to avoid the connections between ground and feed vias. Apart from the mentioned vias, the parallel vias to the feed lines reduce coupling between the feed lines while those near connectors ensure the galvanic connection between the top and bottom substrates. As the PCB is only $0.8$ mm-thick, in order to enhance its mechanic strength and emulate the practical cellphones, $6$- and $3$-mm thick Rohacell-foam was attached to the $28$ and $39$~GHz PCBs, along with $1.5$-mm thick FR4 with grounds. The Rohacell-foam, which has a low permittivity close to the air and a low loss tangent of $0.002$, minimizes losses of the RF signals in the feed lines. In addition, the copper tapes were used to further enhance the electrical connections between the grounds of FR4 and millimeter-wave PCBs. Inside red dotted rectangles of Figs.~\ref{28Phonearrange} and \ref{39Phonearrange} show a cross-sectional view of the whole phone mock-ups. The two mock-ups are different in their widths, antenna array directions and locations, and connector locations. Finally, in order to include the roughness effect of the copper on high-frequency PCBs, $0.4$ $\mu$m roughness of the copper surface \cite{rogers} was applied to the cellphone mock-up simulations.  
\section{Simulation Models of Antenna-Hand Interaction}
\label{sec:simulation}
Electromagnetic simulation models of the antenna-hand interaction are presented in this section. Following a brief introduction to the hand modeling approach, simplified antenna models are provided for feasible antenna-hand simulations. 
\subsection{Photogrammetry Modelling a Human Hand}
\label{sec:Modelling}
The radiation patterns of the cellphone antenna arrays are influenced by the posture of the hand holding the phone. Developing many hand models for antenna-hand interaction simulations is viable. An approach to obtaining 3D human hand models was reported in our prior paper~\cite{pasi2022}. The approach consists of the following 4 steps: 1) taking a video of a hand holding a transparent cellphone-sized box; 2) extracting many pictures from the video and using \textit {Autodesk Recap Photo} to generate a 3D initial hand model; 3) completing the hand model by, e.g., filling a hole, using the software \textit {CloudCompare} and \textit {Autodesk ReCap Photo}, and finally, 4) importing them into electromagnetic (EM) solvers, e.g., \textit {CST Studio Suite}.
\subsection{Simplified Antenna Models}
\begin{figure}[ht] 
    \centering
	\subfigure[]{
	\includegraphics[width=0.48\linewidth]{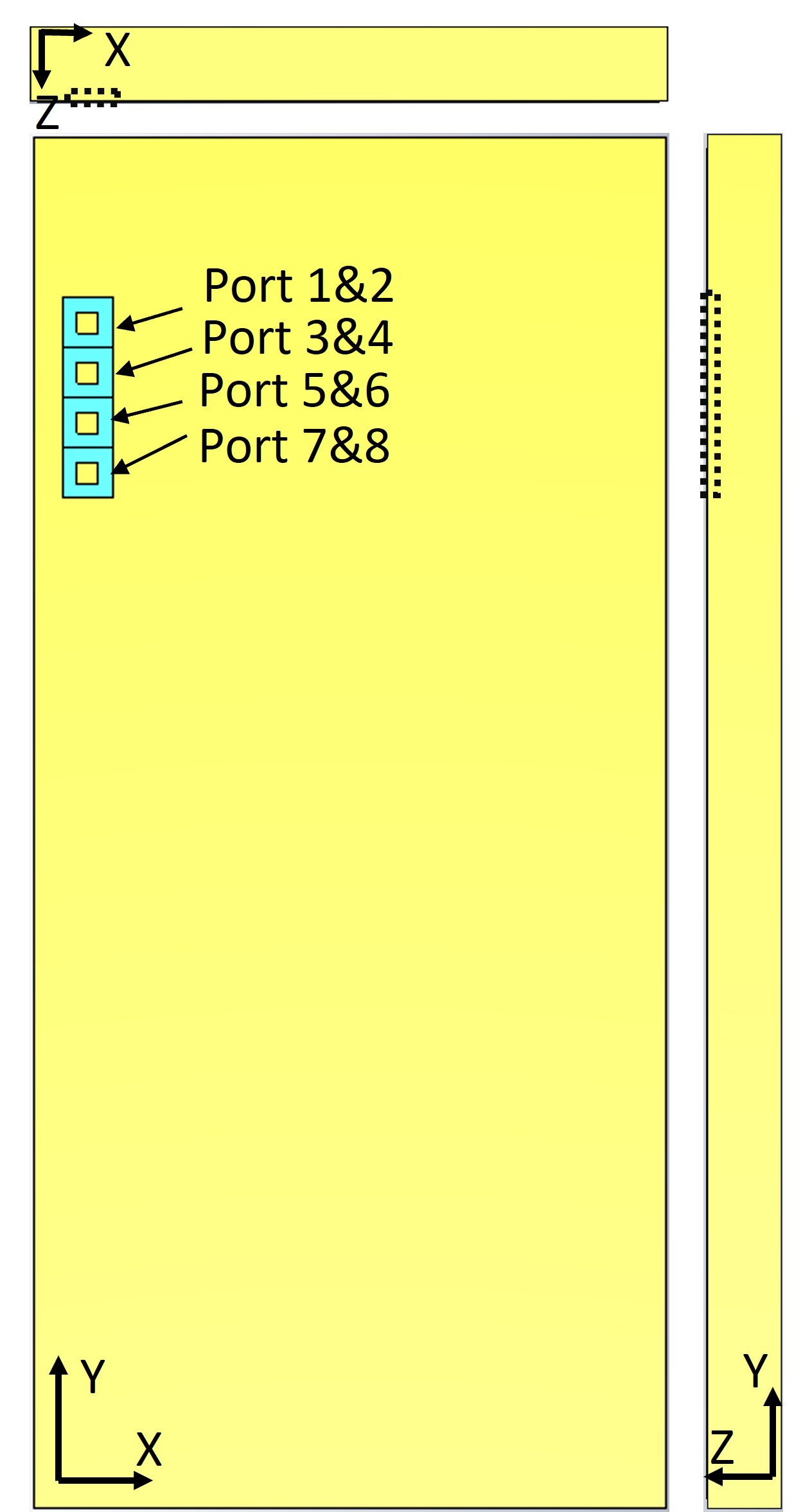}}
	\subfigure[]{
	\includegraphics[width=0.48\linewidth]{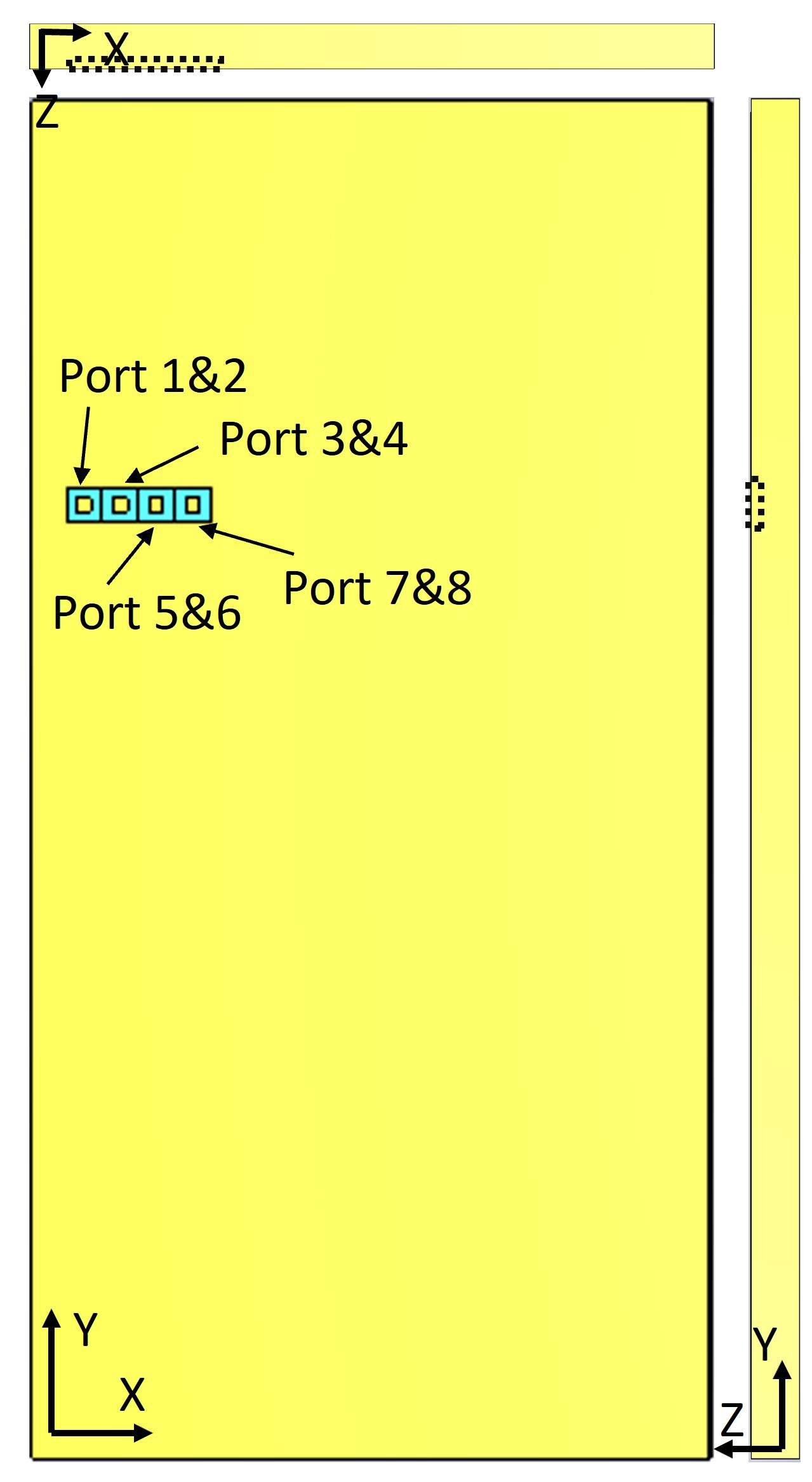}}
	\caption{Schematic of simple antenna array configurations from the top views and the side views of a metal phone chassis (a) at 28 GHz ($69 \times 150 \times 8~{\rm mm^3}$) and (c) at 39 GHz ($75 \times 150  \times 5 {\rm mm^3})$.}
	\label{simplifiedarray}
\end{figure}

Simplified models of cellphone antenna arrays at $28$ and $39$~GHz are illustrated in Fig.~\ref{simplifiedarray}. The simplified arrays consist of patch antennas integrated on a copper cellphone chassis using the same substrates as the complete antenna array models in Figs.~\ref{28Phonearrange} and \ref{39Phonearrange}. The chassis size is the same for the complete and simplified arrays as well as the copper-surface roughness of the antenna arrays. However, the simplified arrays are fed by discrete ports and do not include feed lines or cable connectors. Solving EM simulations with the simplified models is much quicker than those with the complete models. Furthermore, the simplified models are used as a reference to de-embed the losses of the feed lines and cable connectors of the complete array models during measurements. The de-embedding allows us to focus on antenna radiation performance, which is difficult to measure without feed lines or connectors. The cross-verification of the simplified models and the manufactured prototypes is performed in Section~\ref{sec:free-space} by comparing radiation patterns.
\subsection{Antenna-Hand Models}
\begin{figure*}[htbp] 
    \centering
	\includegraphics[width=0.7\linewidth]{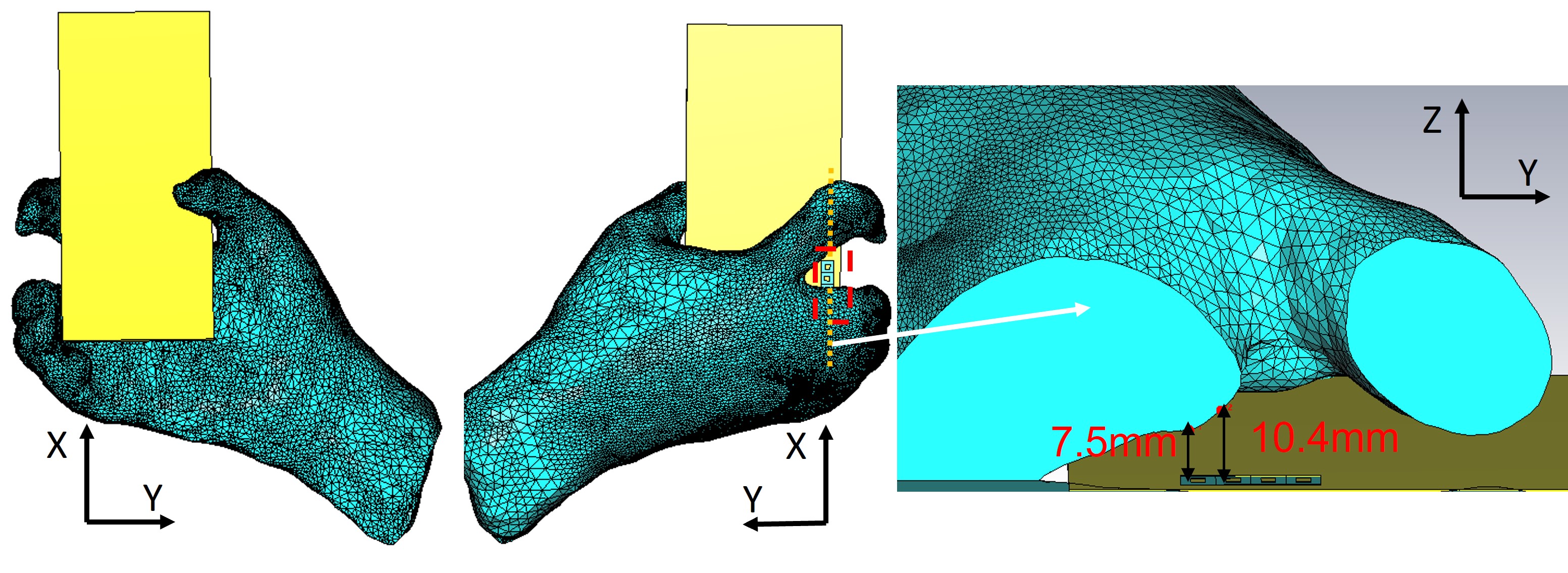}
	\caption{The top and bottom views of the hand with 28 GHz phone mock-up; the cross section along the antenna array with the normal distance between antennas and the hand model.}
	\label{28posture}
\end{figure*}

\begin{figure}[htbp] 
    \centering
	\includegraphics[width=1\linewidth]{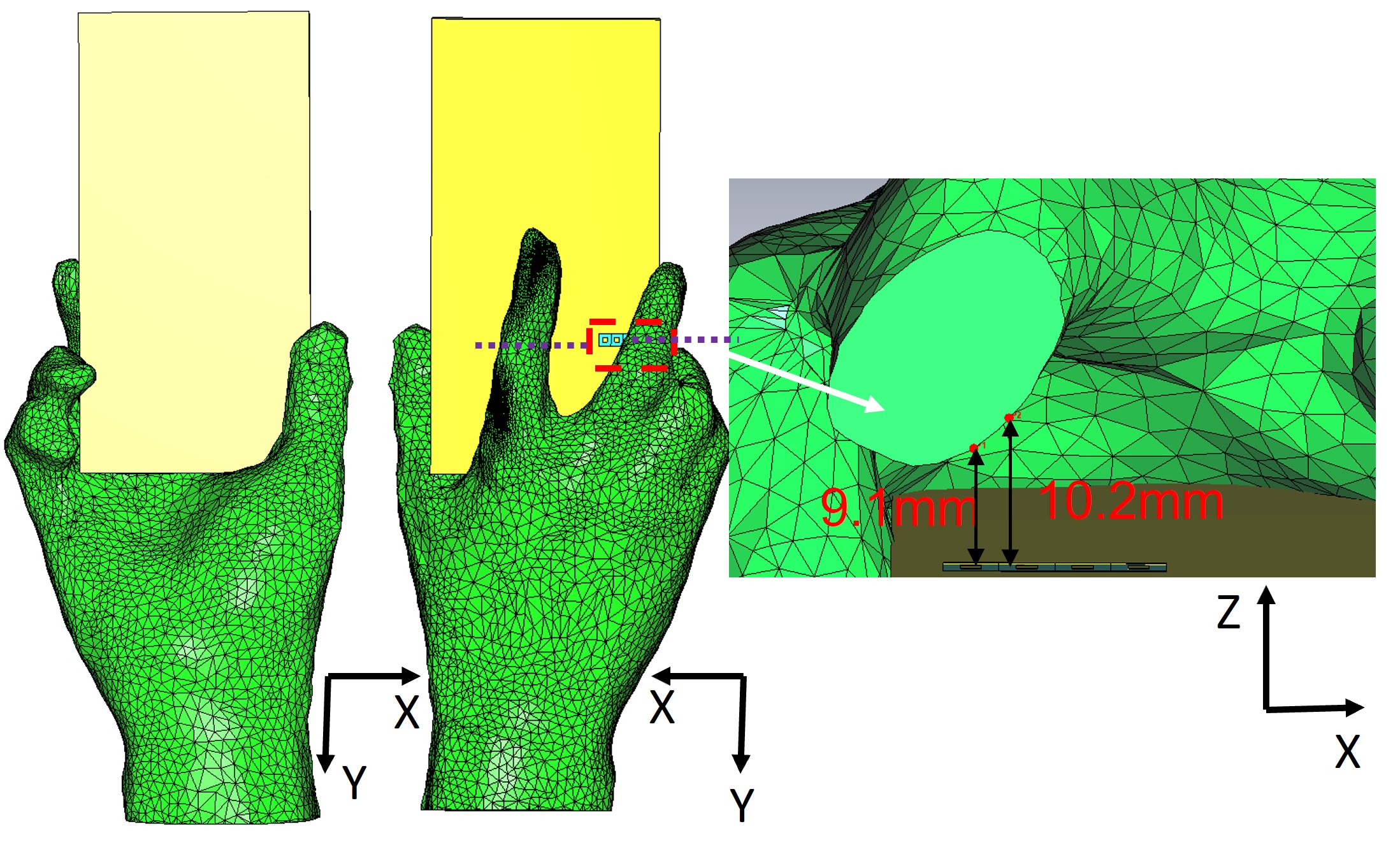}
	\caption{The top and bottom views of the hand with 39 GHz phone mock-up; the cross section along the antenna array with the normal distance between antennas and the hand model.}
	\label{39posture}
\end{figure}

Figs.~\ref{28posture} and~\ref{39posture} show the hand models generated by the approach in \cite{pasi2022} and integrated with the simplified cellphone chassis. They are the {\it exact} postures of hands during antenna array measurements detailed in Section~\ref{sec:antenna-hand}. Some dimensions are also shown on the figure as they are important for the patch antennas not to suffer from impedance matching degradation due to proximity hand tissue. The permittivity of the hands is chosen to be that of dry human skin, i.e., $\epsilon_{\rm r} = 16.55$ and $\sigma = 25.82\  \mathrm{S/m}$ at 28 GHz; $\epsilon_{\rm r} = 11.98$ and $\sigma = 31.43\  \mathrm{S/m}$ at 39 GHz \cite{pasi2022,Gabriel1996}.
\section{Antenna Array Measurements in Free Space}
\label{sec:free-space}
The principle and setups for free-space antenna array measurements are described in this section. Impedance bandwidth, isolation, and radiation characteristics of the manufactured referential arrays are compared with those from simulations.
\subsection{Input Reflection Coefficients and Mutual Coupling}
\begin{figure}[htbp]
    \centering
	\subfigure[]{
	\includegraphics[width=0.8\linewidth]{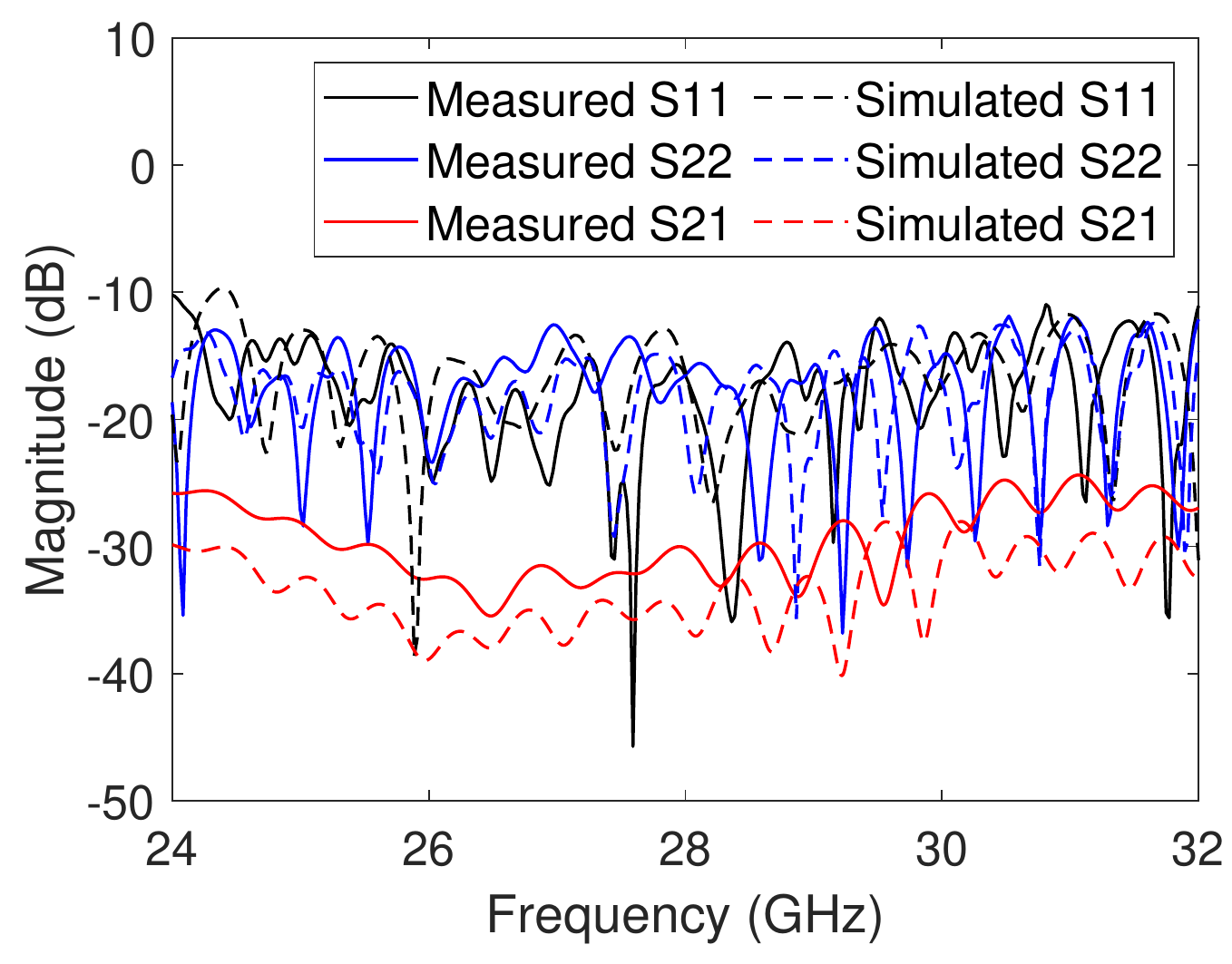}}
	\subfigure[]{
	\includegraphics[width=0.8\linewidth]{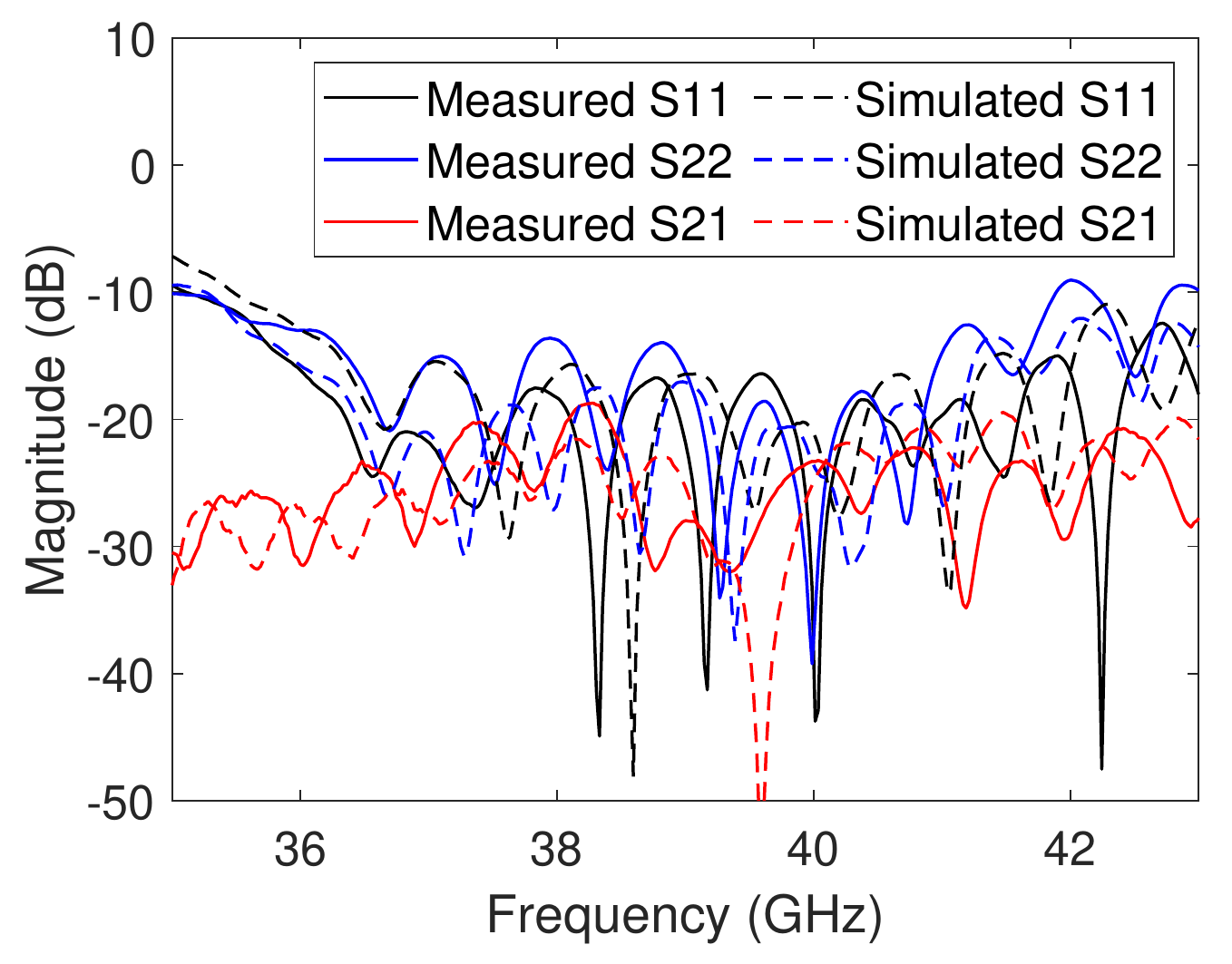}\label{fig:39sparam}}
	\caption{Reflection coefficients and mutual coupling of two exemplary ports 1 and 2 for (a) 28 GHz and (b) 39 GHz cellphone mock-ups. }
	\label{Sparameters}
\end{figure}

The reflection coefficients of the scattering parameter were measured for each port of 28 GHz and 39 GHz phone mock-up by using the PNA-X N5245A vector network analyzer (VNA). The calibration function of the VNA was implemented from 20 GHz to 50 GHz. The bandwidth of intermediate frequency was 1 kHz. Two representative ports were selected to show the impedance matching and the mutual coupling in Fig.~\ref{Sparameters}. The ports' names are already defined in Fig.~\ref{simplifiedarray}. The simulations and measurements show a close agreement although there is an about 400-MHz band shift for the 39 GHz phone mock-up in Fig.~\ref{fig:39sparam}. Still, antennas have wide-enough impedance bandwidth to radiate.
\subsection{Principle of Radiation Measurements}\label{sec:principle}
\begin{figure}[htbp] 
	\centering
	\subfigure[]{
	\includegraphics[width=1\linewidth]{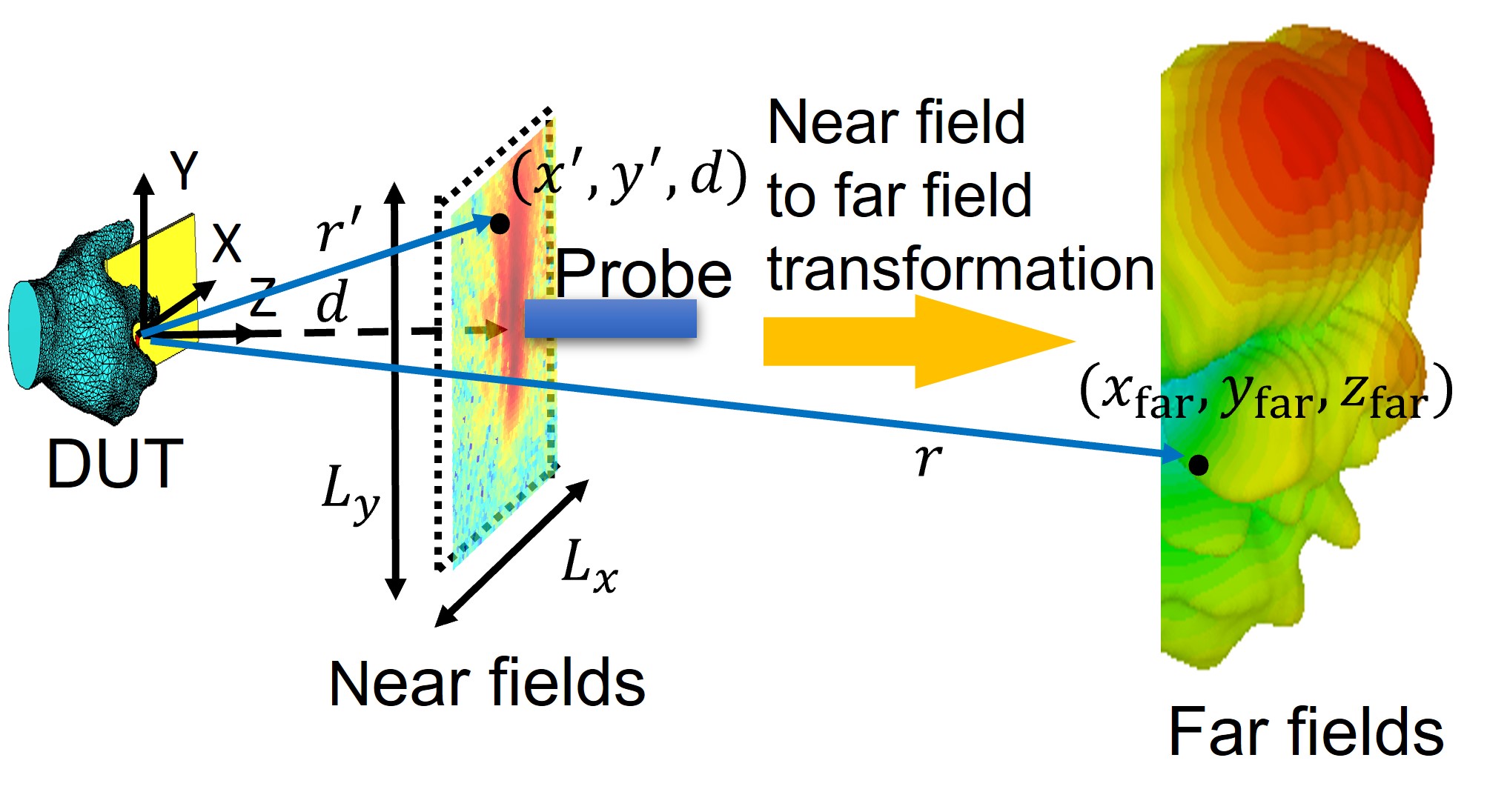}
\label{fig:near_coordinate}}
	\subfigure[]{
	\includegraphics[width=0.7\linewidth]{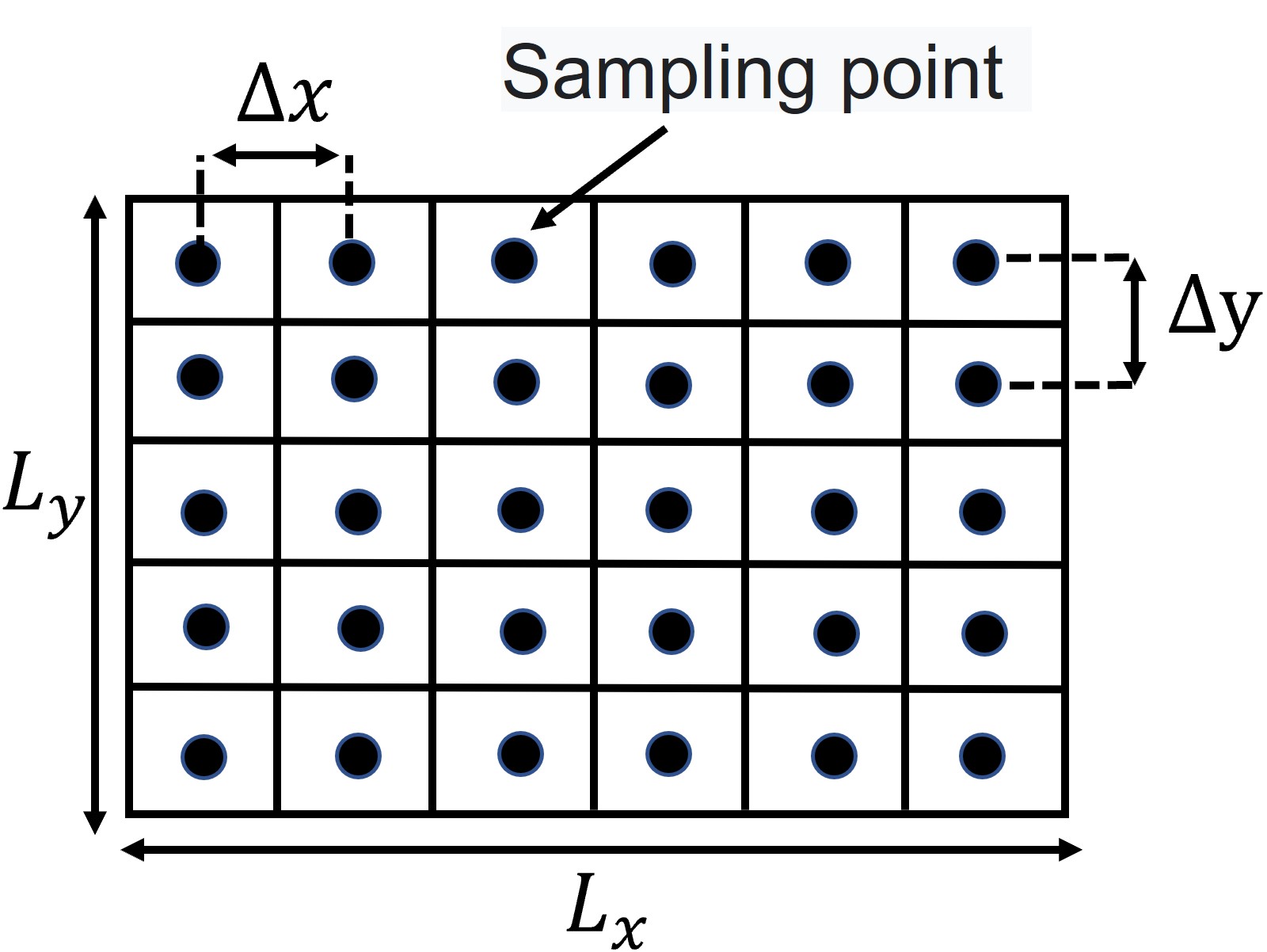}}
	\label{nearmeasurement}
	\subfigure[]{
	\includegraphics[width=0.8\linewidth]{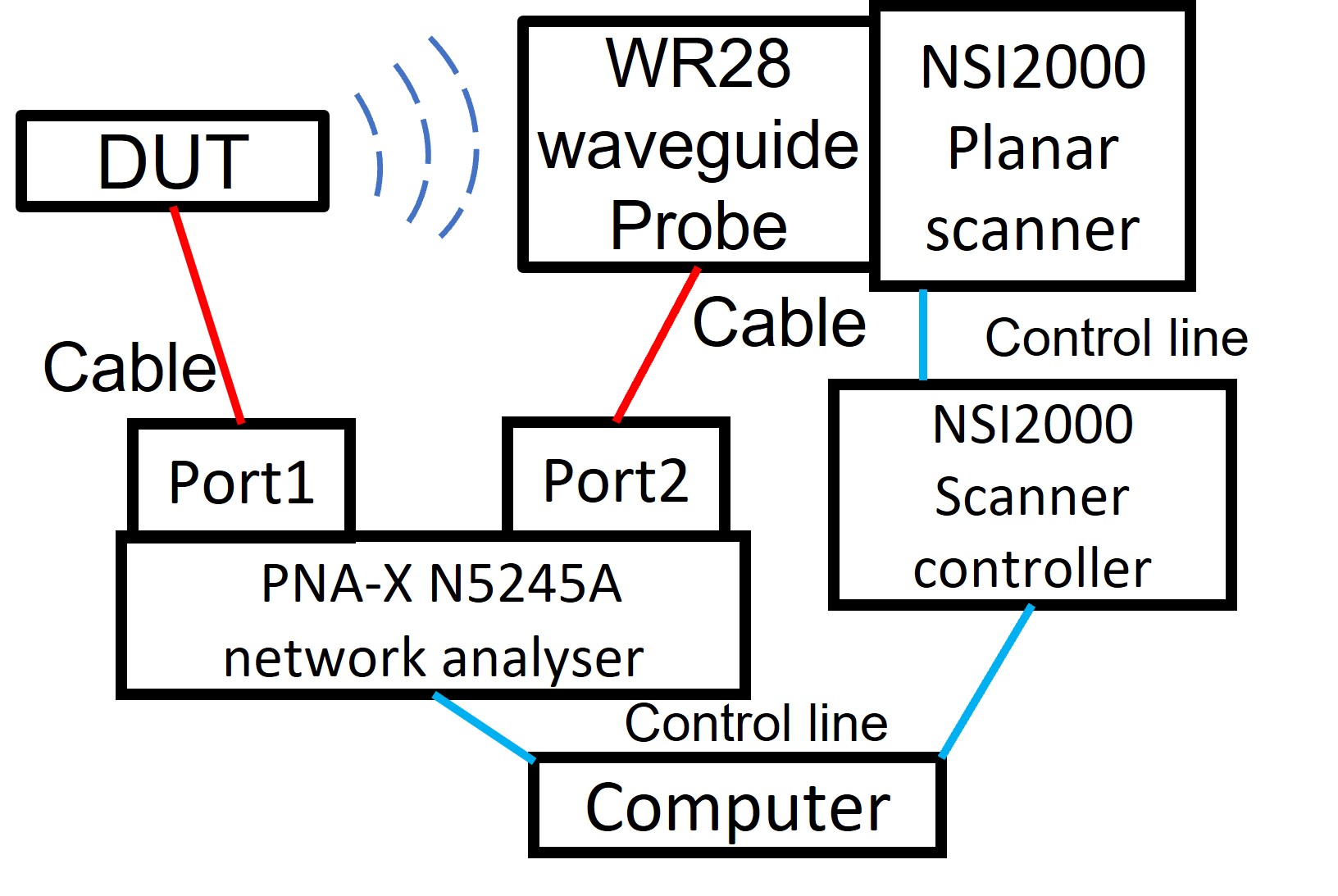}
\label{fig:system}}
	\caption{(a) Definitions of our near-field measurements and far-field observation points, (b) planar probe scanning and (c) schematic of the near-field measurement system for far-field radiation patterns.}
\end{figure}

The schematic of the radiation pattern measurement by planar near-field scanning is shown in Fig.~\ref{fig:near_coordinate}. Spacing between two adjacent near-field sampling points along the $x$- and $y$-axes is $\Delta{x}$ and $\Delta{y}$; $d$ is the normal distance between the antenna array under the test and the probe scanning plane. The measured tangential electric-field components of the near fields on the probe scanning plane are transformed into far fields. 
Based on the Huygens principle, the equivalent sources on the scanning plane can be defined to replace the original source of the antenna array. On a surface $S_{0}$ of an area $L_{x} L_{y}$ in front of the antennas, the equivalent magnetic current sources at a point $(x'_i,y'_j,d)$ can be obtained by an observation of the electric field as
\begin{equation}
\label{eq:M=NE}
{{\boldsymbol{M}(n)} = -\hat{\boldsymbol{n}}\times{\boldsymbol{E}(n)} \ \ \  \mathrm{on}\   S_{0},} 
\end{equation}
where $\hat{\boldsymbol{n}}$ is a unit normal vector of the scanning plane, which corresponds to $\hat{\boldsymbol{z}}$ in Fig.~\ref{fig:near_coordinate}; the symbol $\times$ denotes the outer product; $1 \le i \le N_x$ and $1 \le j \le N_y$ are indices of the probe measurement points along the $x$- and $y$-axes, respectively, and finally, $n = i+(j-1)N_x$, $1 \le n \le N = N_xN_y$ denotes a unique index of probe measurement points. The far field ${\boldsymbol{E}}_{\mathrm{far}}$ can be represented by 
\begin{equation}
{{\boldsymbol{E}}_{\mathrm{far}} = -\nabla \times \int_{S_{0}} {\boldsymbol{M}}(r^{\prime})g(r,r^{\prime})\mathrm{ds}^{\prime},} 
\end{equation}
where $\nabla \times$ is the curl operator in the Cartesian coordinate system and includes the transformation from Cartesian to spherical coordinate systems; ${\boldsymbol{E}}_{\mathrm{far}} = [{E}_{\mathrm{\theta}} \  {E}_{\mathrm{\phi}} \ {E}_{r}]^\top$ and ${\boldsymbol{M}} = [{M}_{x} \ {M}_{y} \ 0]^\top$, where ${E}_{r} \approx 0$ when $r \rightarrow \infty$ and $\cdot^\top$ represents a transpose operation; $g(r,r^{\prime})=\frac{e^{-\mathrm{j} k_{0}\left|{r}-{r}^{\prime}\right|}}  {4\pi\left|{r}-{r}^{\prime}\right|}$ is the Green's function in free space; $r’ = \sqrt{(x’)^2 + (y’)^2 + d^2} $ is a distance from the origin of the coordinate system to the probe measurement point $(x’, y’, d)$; $r = \sqrt{(x_{\rm far})^2 + (y_{\rm far})^2 + (z_{\rm far})^2} $ is a distance from the origin of the coordinate system to the far-field point $(x_{\rm far}, y_{\rm far}, z_{\rm far})$ as defined in Fig.~\ref{fig:near_coordinate}; $k_{0}$ is the wave number in free space at the testing frequency. Though the origin can be set arbitrary, in this paper, it is located at the middle of the antenna array. Then, the $x$- and $y$-components of the equivalent magnetic current source $\boldsymbol{M}$ in~\eqref{eq:M=NE} can be collected across $N$ near-field measurement points to form column vectors as 
\begin{eqnarray}
\label{eq:Mx}
\boldsymbol{M}_{x} & = & \left[ E_y(1)~ E_y(2)~\cdots~ E_y(N)~ \right]^\top, \\
\boldsymbol{M}_{y} & = & -\left[ E_x(1)~ E_x(2)~\cdots~ E_x(N)~ \right]^\top.
\end{eqnarray}
Then, the $\theta$- and $\phi$-components of the far-field ${\boldsymbol{E}}_{\mathrm{far}}$ observed at $(x_{\mathrm{far},k}$, $y_{\mathrm{far},k}, z_{\mathrm{far},k})$, $1 \le k \le K$, can be obtained by 
\begin{equation}
\begin{bmatrix}
    E_{\mathrm{\theta},k}  \\
    E_{\mathrm{\phi},k}
\end{bmatrix} = 
\begin{bmatrix}
    \boldsymbol{H}_{k,1 1} & 
    \boldsymbol{H}_{k,1 2} \\
    \boldsymbol{H}_{k,2 1} & 
    \boldsymbol{H}_{k,2 2}
\end{bmatrix}
\begin{bmatrix}
    \boldsymbol{M}_{x} \\
     \boldsymbol{M}_{y}
\end{bmatrix},
\end{equation}
where $\boldsymbol{H}_{k,1 1}$, $\boldsymbol{H}_{k,1 2}$, $\boldsymbol{H}_{k,2 1}$, $\boldsymbol{H}_{k,2 2}$ $\in \mathbb{C}^{N}$ are row vectors. Their $n$-th entry, denoted as $[ \boldsymbol\cdot ]_n$, can be represented respectively by
\begin{equation}
\begin{bmatrix}
    \boldsymbol{H}_{k,1 1}
\end{bmatrix}_n = 
\begin{Bmatrix}
\cos{\theta_k}\sin{\phi_k}(z_{\mathrm{far},k} - d) \\ 
+ \sin{\theta_k}(y_{\mathrm{far},k} - y'_{n})
\end{Bmatrix}G^{\prime}(R_{k,n})\Delta{x}\Delta{y} 
\end{equation}
\begin{equation}
\begin{bmatrix}
    \boldsymbol{H}_{k,1 2}
\end{bmatrix}_n = - 
\begin{Bmatrix}
\cos{\theta_k}\cos{\phi_k}(z_{\mathrm{far},k} - d) \\ 
+ \sin{\theta_k}(x_{\mathrm{far},k} - x'_{n})
\end{Bmatrix}G^{\prime}(R_{k,n})\Delta{x}\Delta{y} 
\end{equation}
\begin{equation}
\begin{bmatrix}
    \boldsymbol{H}_{k,2 1}
\end{bmatrix}_n = 
\begin{Bmatrix}
\cos{\phi_k}(z_{\mathrm{far},k} - d)
\end{Bmatrix}G^{\prime}(R_{k,n})\Delta{x}\Delta{y} 
\end{equation}
\begin{equation}
\begin{bmatrix}
    \boldsymbol{H}_{k,2 2}
\end{bmatrix}_n = 
\begin{Bmatrix}
\sin{\phi_k}(z_{\mathrm{far},k} - d)
\end{Bmatrix}G^{\prime}(R_{k,n})\Delta{x}\Delta{y},
\end{equation}
where $G^{\prime}(R_{k,n}) = \frac{e^{-\mathrm{j} k_{0}R_{k,n}}}{4\pi R_{k,n}}(\mathrm{j} k_{0} + \frac{1}{R_{k,n}})$ and $R_{k,n}$ is the distance between the $k$-th far-field point and the $n$-th probe measurement point. By using the equations, the radiation patterns of the antenna under the test can be obtained based on the planar near-field measurements.

\subsection{Free Space Measurements}
Free space measurements are performed to verify the near-field to far-field transformation method and to estimate losses of the measurement system for de-embedding, serving for calibration. 

The schematic of the near-field scanner system is shown in Fig.~\ref{fig:system}. A standard rectangular WR28 waveguide was used as the field probe installed on the NSI2000 planar scanner. The PNA-X N5245A network analyzer was used to measure the transmission coefficients of the scattering parameter between the antenna under test and the field probe. A computer controlled positioning of the planar scanner for the waveguide probe along with acquisition of the transmission coefficients. The bandwidth of intermediate frequency in the VNA was $10$~kHz, while its output power was $5$~dBm. When measuring the two polarizations $E_x$ and $E_y$, it was necessary to repeat the scanning for each polarization after changing the waveguide orientation by $90^\circ$.

In practice, the fast and accurate near-field measurements should meet some requirements:
\begin{enumerate}{}{}
\item{the normal-direction spacing $d$ between antenna elements and the probe should be 1 $\sim$ 5 wavelengths of the frequency of interest;}
\item{the measurement steps $\Delta{x},\Delta{y}$ along $x$- and $y$-axes should be smaller than 0.5 wavelengths;}
\item{the limited scanning area $S_{0}$ is chosen as long as the power density of the radiated field from the antenna under test is small enough outside the area and the signal-noise ratio (SNR) of the received signal at the VNA is high. However, the low power density outside the area $S_{0}$ and high SNR cannot be always achieved simultaneously. Accordingly, a trade-off between the power density and SNR is considered. }
\end{enumerate}
The setups of 28 GHz and 39 GHz near-field scanning are defined based on the measurement requirements as shown in TABLE \ref{setups1}. 
\begin{table}[htbp]
	\begin{center}
		\caption{SETUPS OF NEAR FIELD SCANNING FOR CELLPHONE MOCK-UPS IN FREE SPACE AND WITH HAND EFFECTS} 
		\label{setups1}
		\vspace{11pt}
		\begin{tabular}{|c|c|c|c|c|}\hline &\multicolumn{2}{c|}{Free Space} & \multicolumn{2}{c|}{With Hand}\\ \hline 
		Frequency [GHz] &28 & 39 &28 & 39 \\ \hline
		$d$ [mm]         &$20$ &$20$ &$50$ &$50$ \\ \hline 
	    $\Delta{x}$ [mm] &$5$ &$4$ &$5$ &$4$ \\ \hline 
		$\Delta{y}$ [mm] &$5$ &$4$ &$5$ &$4$ \\ \hline 
		$L_x$ [mm]       &$400$ &$300$ &$200$ &$200$ \\ \hline 
		$L_y$ [mm]       &$400$ &$300$ &$200$ &$200$\\ \hline
		Measurement time &\multirow{2}{*}{100}&\multirow{2}{*}{80} &\multirow{2}{*}{12}&\multirow{2}{*}{15} \\
		for a single port [min] &&&&\\ \hline
		Only major &\multirow{2}{*}{No}&\multirow{2}{*}{No}&\multirow{2}{*}{Yes}&\multirow{2}{*}{Yes} \\ 
		polarization &&&& \\ \hline
 		\end{tabular}
	\end{center}
	\end{table}
\subsection{De-embedding the Losses of the Measurement System}\label{sec:deembedding}
During the near-field to far-field transformation, the far-field locations are two meters away from the antenna array, i.e., $r=2$ m. As shown in Fig.~\ref{28Phonearrange} and Fig.~\ref{39Phonearrange}, the manufactured antenna arrays are with the different lengths of the feed lines and cable connectors. In order to evaluate the characteristics of the antenna array only, the losses of ancillary parts of the antenna array and due to the measurement system, including the near- to far-field transformation method, need to be de-embedded. Their complex-valued losses are estimated by the average difference between the simulated and measured main beam in free space~\cite{Lauri2021}. Numerically simulated far fields of the simplified antenna models in Fig.~\ref{simplifiedarray} are used as the reference in estimating the losses, while the measurements cover $\phi \in [0\degree,360\degree]$ and $\theta\in [0\degree,60\degree]$ since the far fields are only valid for $\theta \le 60\degree$ in the near-field scanner, refer to Fig.~\ref{fig:near_coordinate} for the coordinate system. The loss estimates of each port for the two prototypes are shown in TABLE \ref{deembedloss2}, indicating that they are proportional to the lengths of the feed lines. 

\begin{figure}[htbp] 
    \centering
	\subfigure[]{
	\includegraphics[width=0.8\linewidth]{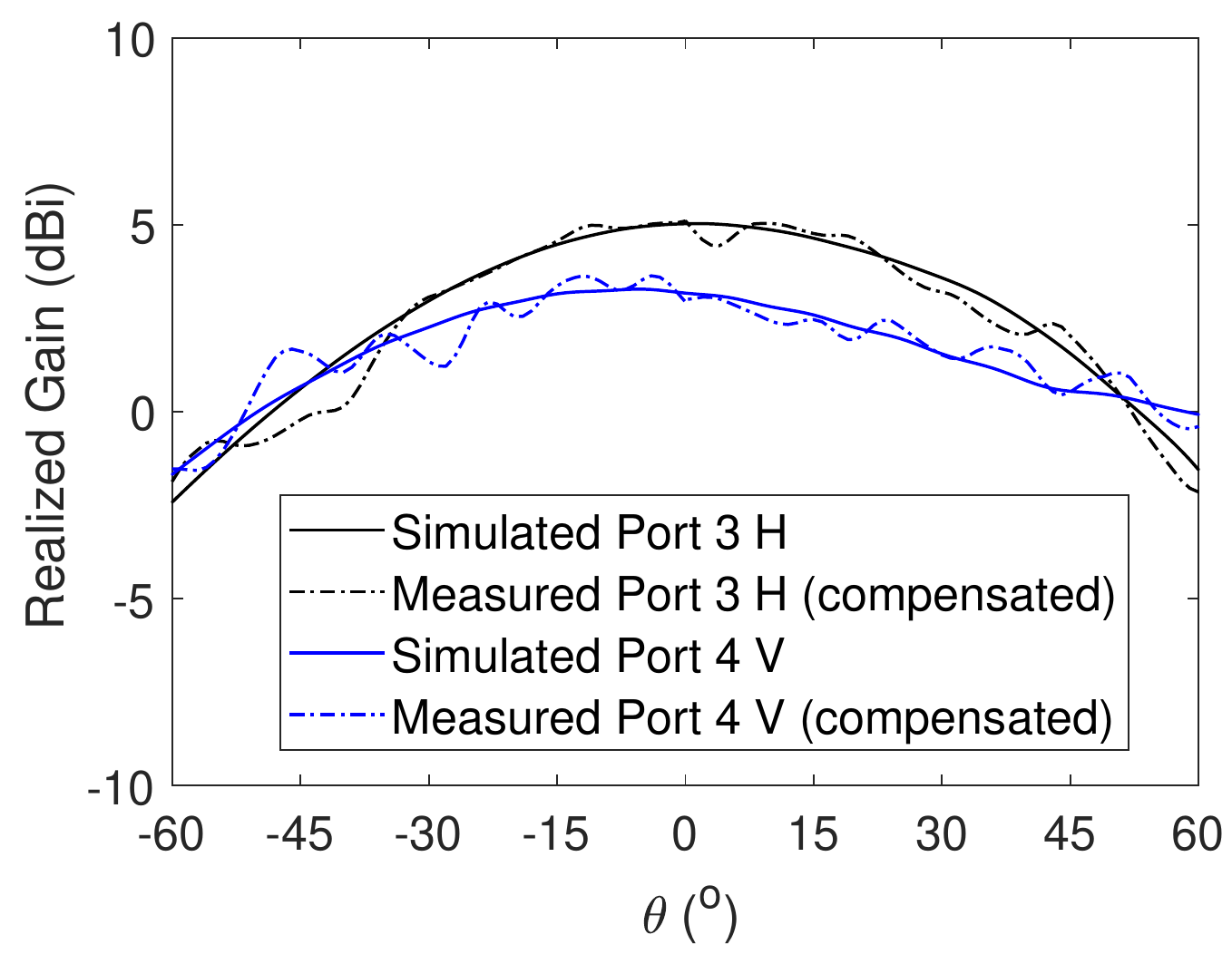}}
	\subfigure[]{
	\includegraphics[width=0.8\linewidth]{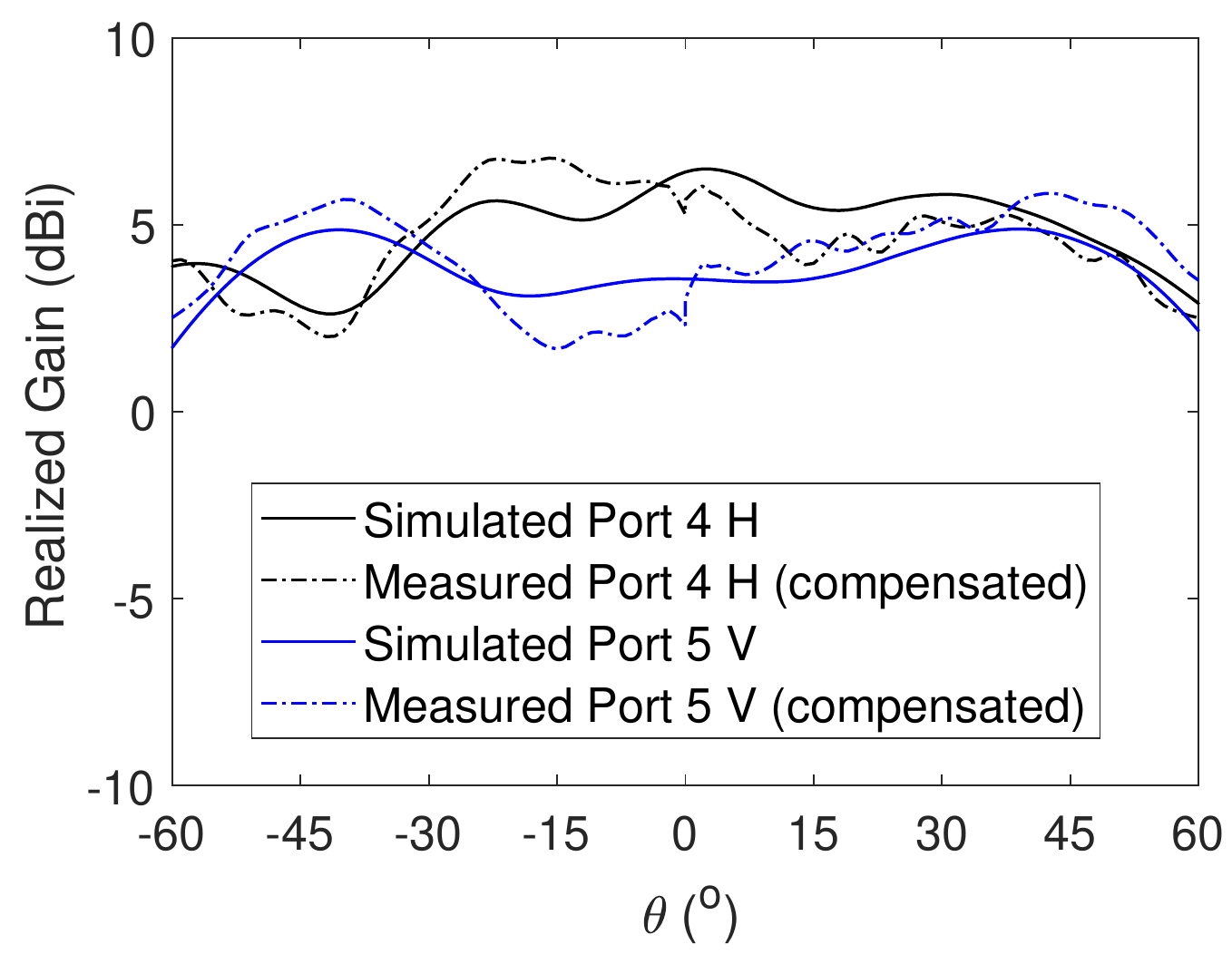}}
	\caption{Elevation cuts of the realized far-field gains in free space for (a) 28 GHz mock-up (Port 3 H and Port 4 V) and (b) 39 GHz mock-up (Port 4 H and Port 5 V) obtained by planner near-field scanning.}
	\label{nearfreespaceGain}
\end{figure}

After de-embedding the losses, the elevation cuts of the realized gains in free space are shown in Fig.~\ref{nearfreespaceGain} for the selected ports of the two cellphone mock-ups. 'H' or 'V' shows the vertical or horizontal polarization of ports based on the coordinates in Fig.~\ref{simplifiedarray}. The simplified models' radiation patterns are similar to those of the manufactured prototypes where the maximum difference of $1.5$~dB, indicating that the simplified models are good representative models of the manufactured prototypes. 
\begin{table}[htbp]
	\begin{center}
		\caption{ESTIMATED LOSSES OF EACH PORT OF THE ARRAYS DUE TO THE ANCILLARY PARTS OF THE MANUFACTURED ARRAY AND THE MEASUREMENT SYSTEM INCLUDING THE NEAR- TO FAR-FIELD TRANSFORMATION METHOD} \label{deembedloss2}
		\vspace{11pt}
		\begin{tabular}{|c|c|c|}\hline Ports &28 GHz & 39 GHz  \\ \hline
		1 &$93.0$ dB&$99.1$ dB\\ \hline 
	    2 &$91.3$ dB&$97.3$ dB \\ \hline 
		3 &$91.8$ dB&$97.0$ dB\\ \hline
    	4 &$90.9$ dB&$95.3$ dB\\ \hline
		5 &$91.5$ dB&$96.9$ dB\\ \hline
		6 &$91.1$ dB&$97.2$ dB\\ \hline
	    7 &$90.8$ dB&$97.3$ dB\\ \hline 
	    8 &$91.0$ dB&$97.4$ dB\\ \hline 
 		\end{tabular}
	\end{center}
	\end{table}
The loss estimates are also applied to the far-field patterns derived from the near-field measurements, both with and without hand effects. In the latter, the detuning effects of real hands on the antenna arrays are avoided as will be elaborated in Section~\ref{sec:antenna-hand}.
\subsection{Spherical Coverage}
\label{sec:spherical_coverage}
\begin{figure}[htbp] 
    \centering
	\subfigure[]{
	\includegraphics[width=1\linewidth]{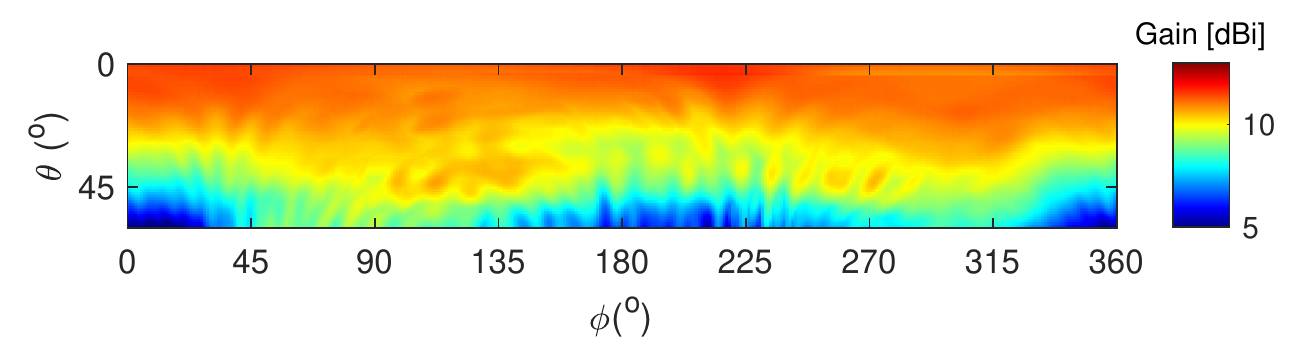}}
	\subfigure[]{
	\includegraphics[width=1\linewidth]{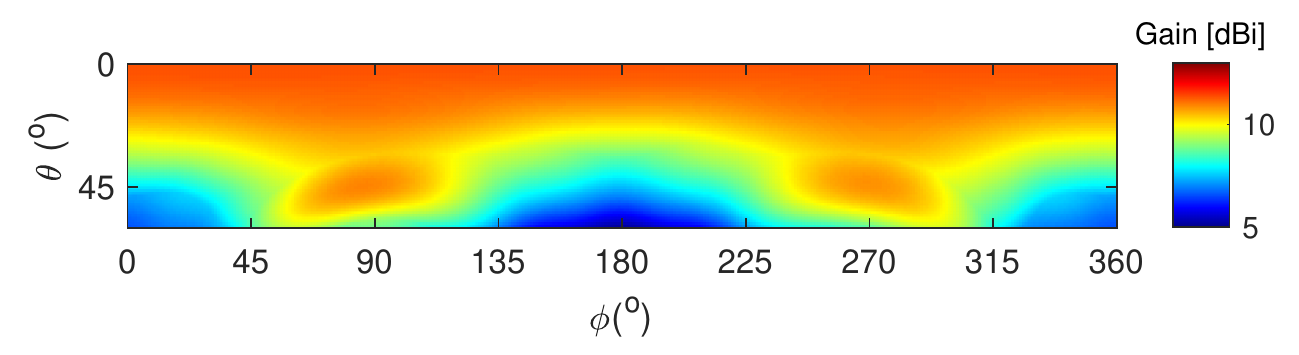}}
	\caption{Spherical coverage of the realized gains in free space for 28 GHz mock-up. (a) Measurement and (b) simulation.}
	\label{28SCnearfreeGain}
\end{figure}

\begin{figure}[htbp] 
    \centering
	\subfigure[]{
	\includegraphics[width=1\linewidth]{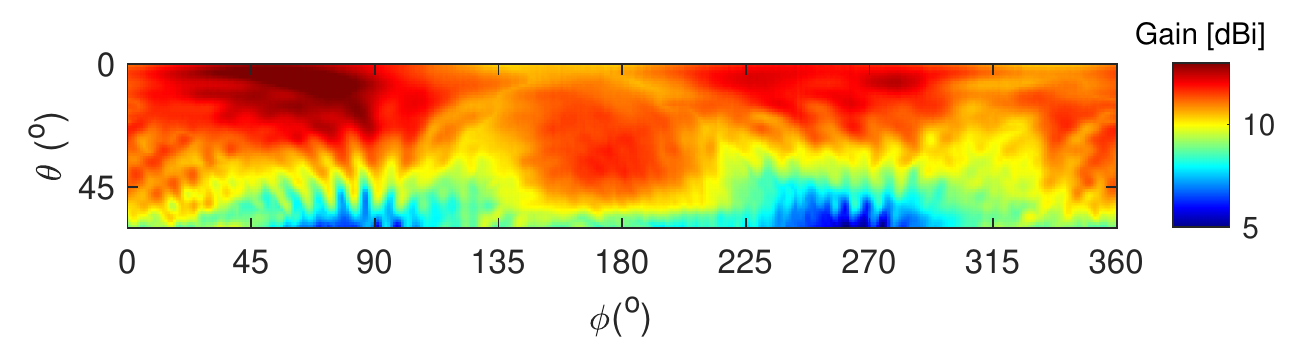}}
	\subfigure[]{
	\includegraphics[width=1\linewidth]{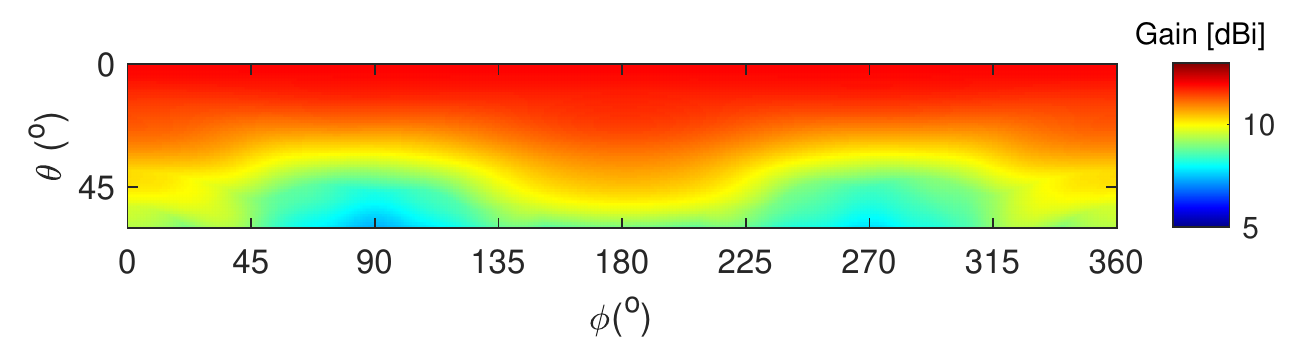}}
	\caption{Spherical coverage of the realized gains in free space for 39 GHz mock-up. (a) Measurement and (b) simulation.}
	\label{39SCnearfreeGain}
\end{figure}

\begin{figure}[htbp] 
    \centering
	\includegraphics[width=0.8\linewidth]{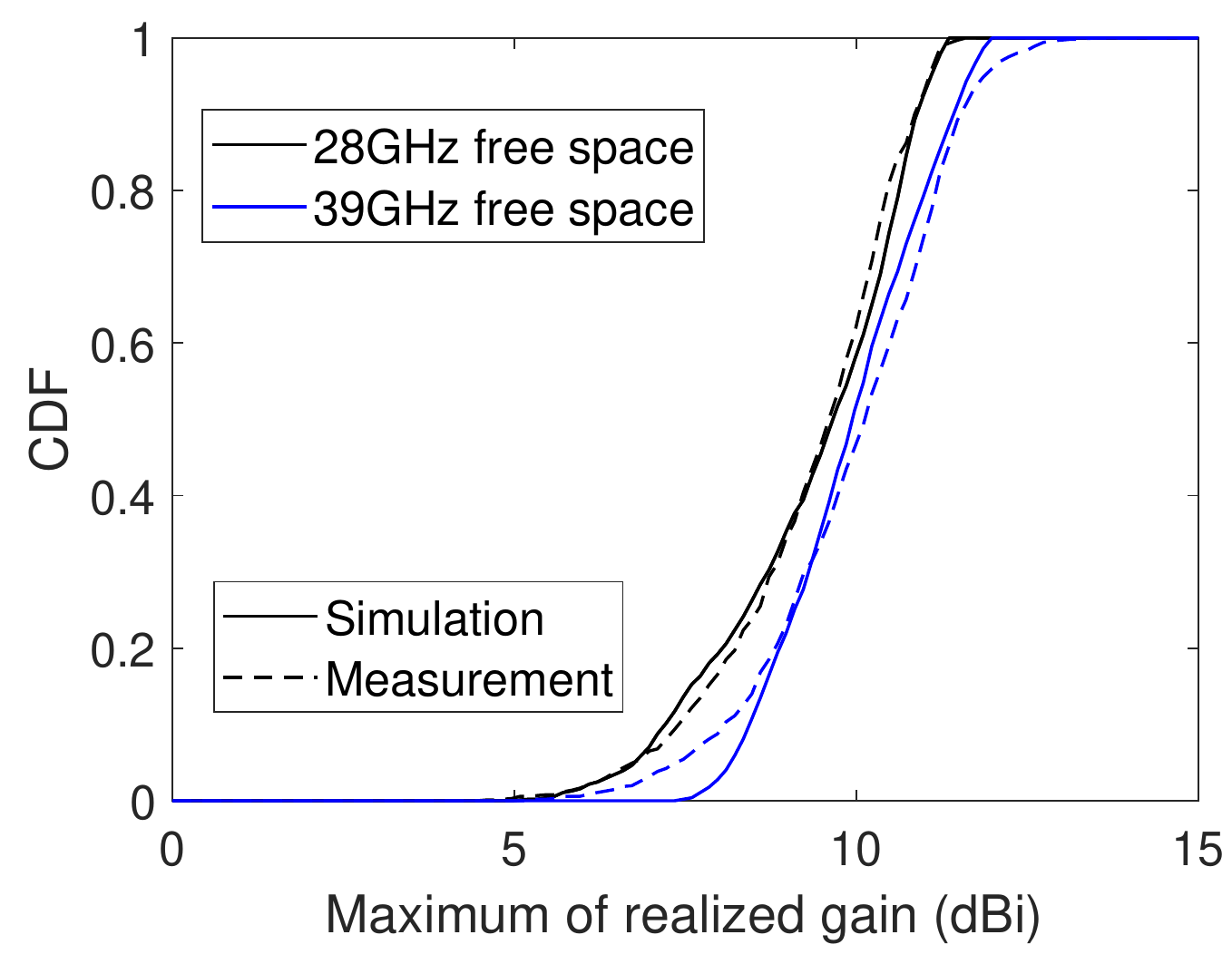}
	\caption{The CDF of spherical coverage for the free-space case.}
	\label{CDFfreespcae}
\end{figure}

Spherical coverage is an empirical statistic of the maximum gains that an antenna array can realize for all feasible angles on a sphere, hence has been one of the important figure-of-merits of millimeter-wave antenna arrays~\cite{Lauri2021, Hazmi2019, Hazmi2020}. When synthesizing a pattern of the array from those of individual antenna ports, equal gain combining is used for each polarization to define complex weights assuming that a single plane wave is incident. For a specific angle $\mathbf\Omega = (\theta,\phi)$ where $\theta$ and $\phi$ denote the polar and azimuth angles respectively, the realized gain of the array after the equal gain combining is given by $\hat{G}(\mathbf{\Omega}) = \sqrt{ | E_{\theta}(\mathbf{\Omega}) |^2 + | E_{\phi}(\mathbf{\Omega})|^2 }$ where $ E_{\theta}(\mathbf{\Omega})$ and $E_{\phi}(\mathbf{\Omega})$ are the complex electric field vectors after the array synthesis for $\theta$- and $\phi$-polarized fields. Then its cumulative plot can be defined by 
\begin{equation}
{CDF(g) = \mathrm{prob}(\hat{G}(\mathbf{\Omega}) <g),}
\label{eq:cdf}
\end{equation}
where $\rm prob(\cdot)$ is a probability operator yielding values between $0$ and $1$. When implementing the spherical coverage statistics, any chosen angles $\mathbf\Omega$ must be uniform on the whole sphere, which means the number of azimuth angle samples is smaller at the higher elevation angles. The uniform grid over the whole sphere can ensure that adjacent points are with the same angular distance~\cite{3GPP2017}. In this paper, 4000 points are chosen to be uniform over the valid angle range ($\phi \in [0\degree,360\degree]$ and $\theta\in [0\degree,60\degree]$). The spherical coverage statistics are valid even if the antenna array cannot steer beams to the whole angular range.
Spherical coverage of 28 GHz and 39 GHz cellphone mock-ups is shown in Figs.~\ref{28SCnearfreeGain} and \ref{39SCnearfreeGain}. The gain distributions are quite similar although some ripples can be seen in the measurement results. The cumulative distribution function (CDF) of the spherical coverage is computed using~\eqref{eq:cdf}. In Fig.~\ref{CDFfreespcae}, the gain differences for the 28 GHz cellphone mock-up are smaller than 0.5 dB. For the 39 GHz cellphone mock-up, the gain differences are observed at $CDF < 0.15$ and $CDF > 0.95$, while the difference is below 0.5 dB at the median level. The mentioned comparisons between simulations and measurements show the validity of the simplified antenna array model and the array measurement method through near- to far-field transformation and loss de-embedding. 
\section{Measurements of Antenna-hand Interaction}
\label{sec:antenna-hand}
Having established the radiation pattern measurement method in free space, in this section, our approaches of real-hand measurements and their results are shown.
\begin{figure*}[htbp] 
    \centering
    \subfigure[]{
	\includegraphics[width=0.59\linewidth]{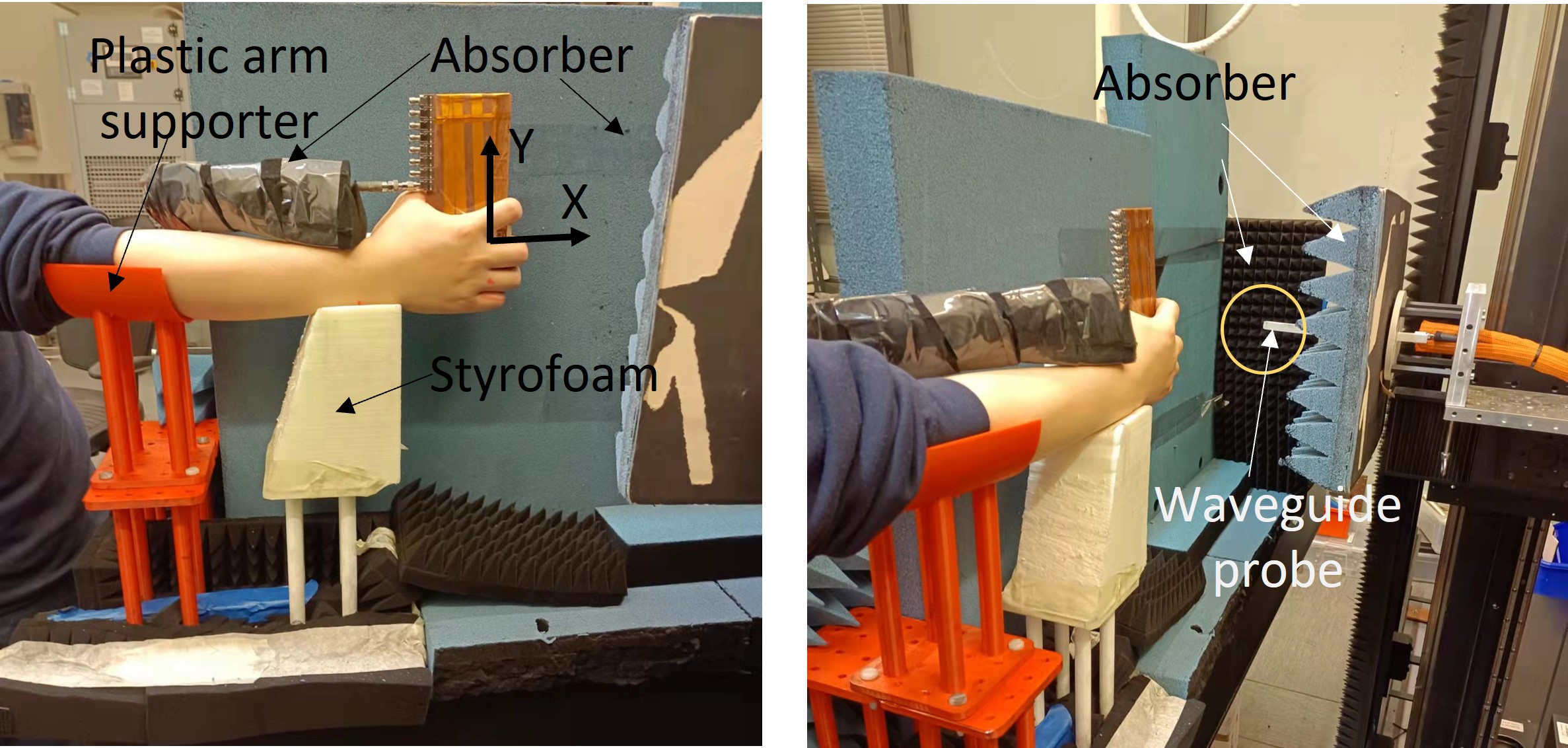}\label{28setups}}
	\subfigure[]{
	\includegraphics[width=0.375\linewidth]{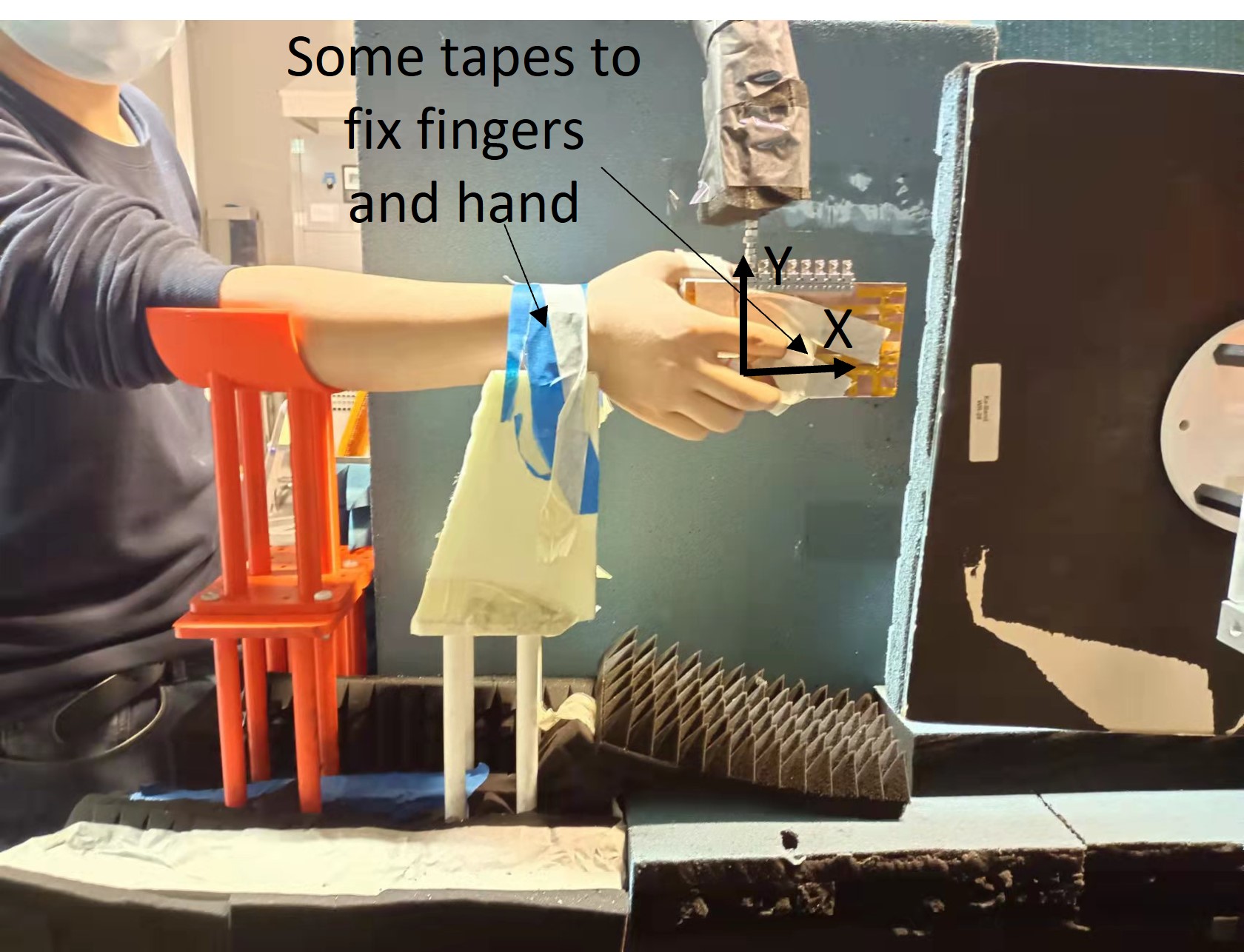}
	\label{39setups}}
	\caption{(a) Two different views of the measurement setups for the 28 GHz antenna array with a human operator. (b) The view of the measurement setups for the 39 GHz antenna array with a human operator.}
\end{figure*}
\begin{figure}[htbp]
    \centering
	\subfigure[]{
	\includegraphics[width=1\linewidth]{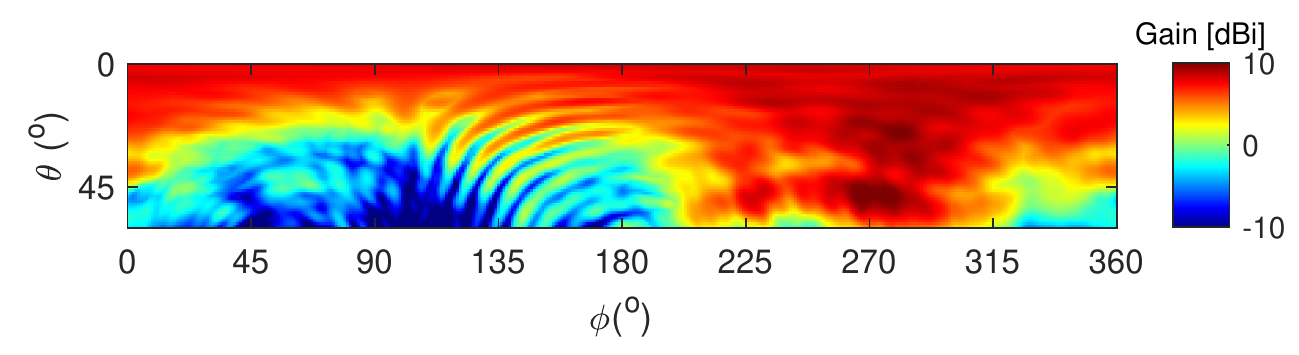} \label{fig:28_meas1}}
	\subfigure[]{
	\includegraphics[width=1\linewidth]{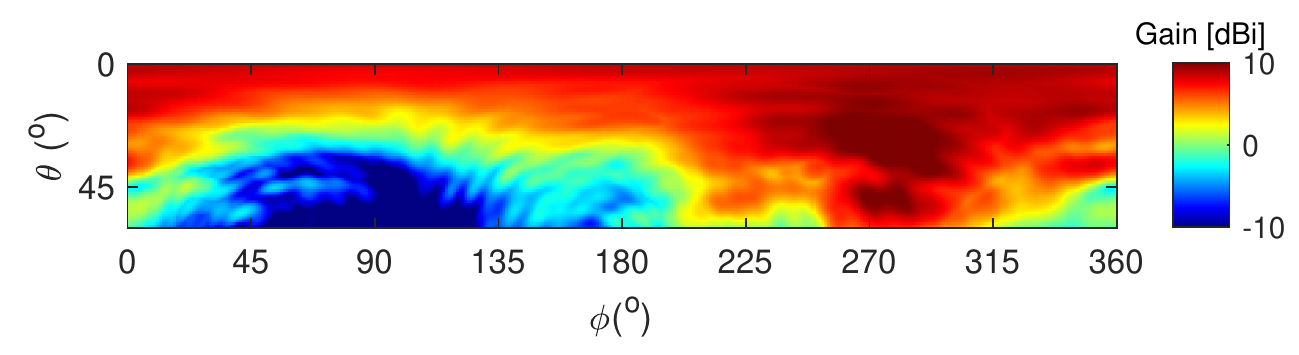} \label{fig:28_meas2}}
	\subfigure[]{
	\includegraphics[width=1\linewidth]{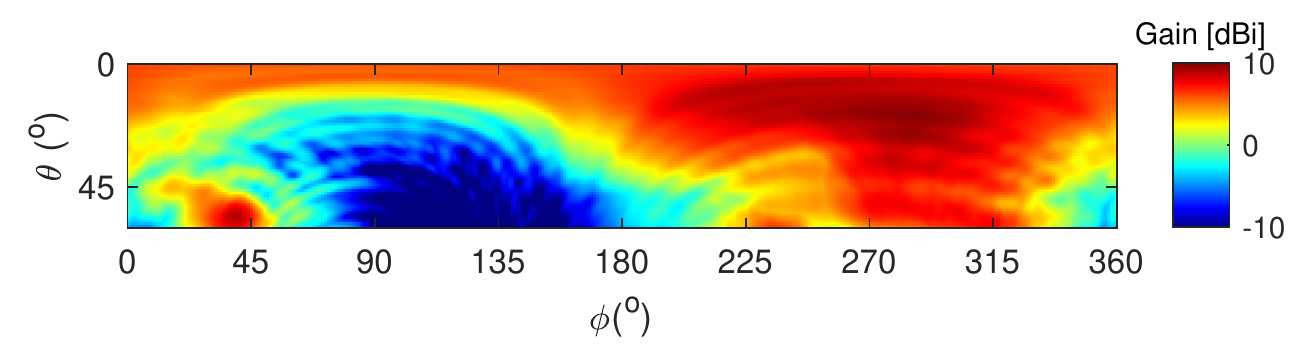} \label{fig:28_simNF}}
	\subfigure[]{
	\includegraphics[width=1\linewidth]{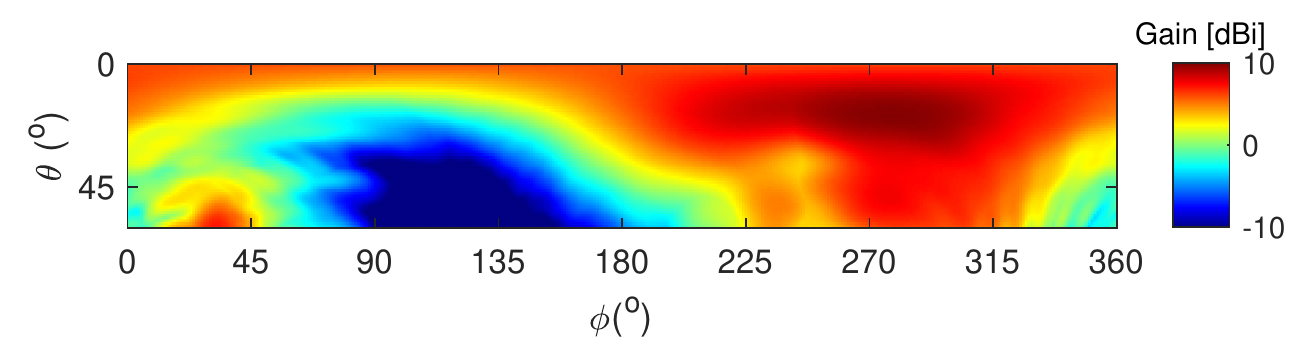} \label{fig:28_simFF}}
	\caption{Spherical coverage of the realized gains of the 28GHz phone mock-up when held by an operator.  (a) The first measurement, (b) the second measurement, (c) near-field simulation with transformation and (d) far-field simulation. }
	\label{28handsphericalcoverage}
\end{figure}
\begin{figure}[ht] 
    \centering
	\includegraphics[width=0.8\linewidth]{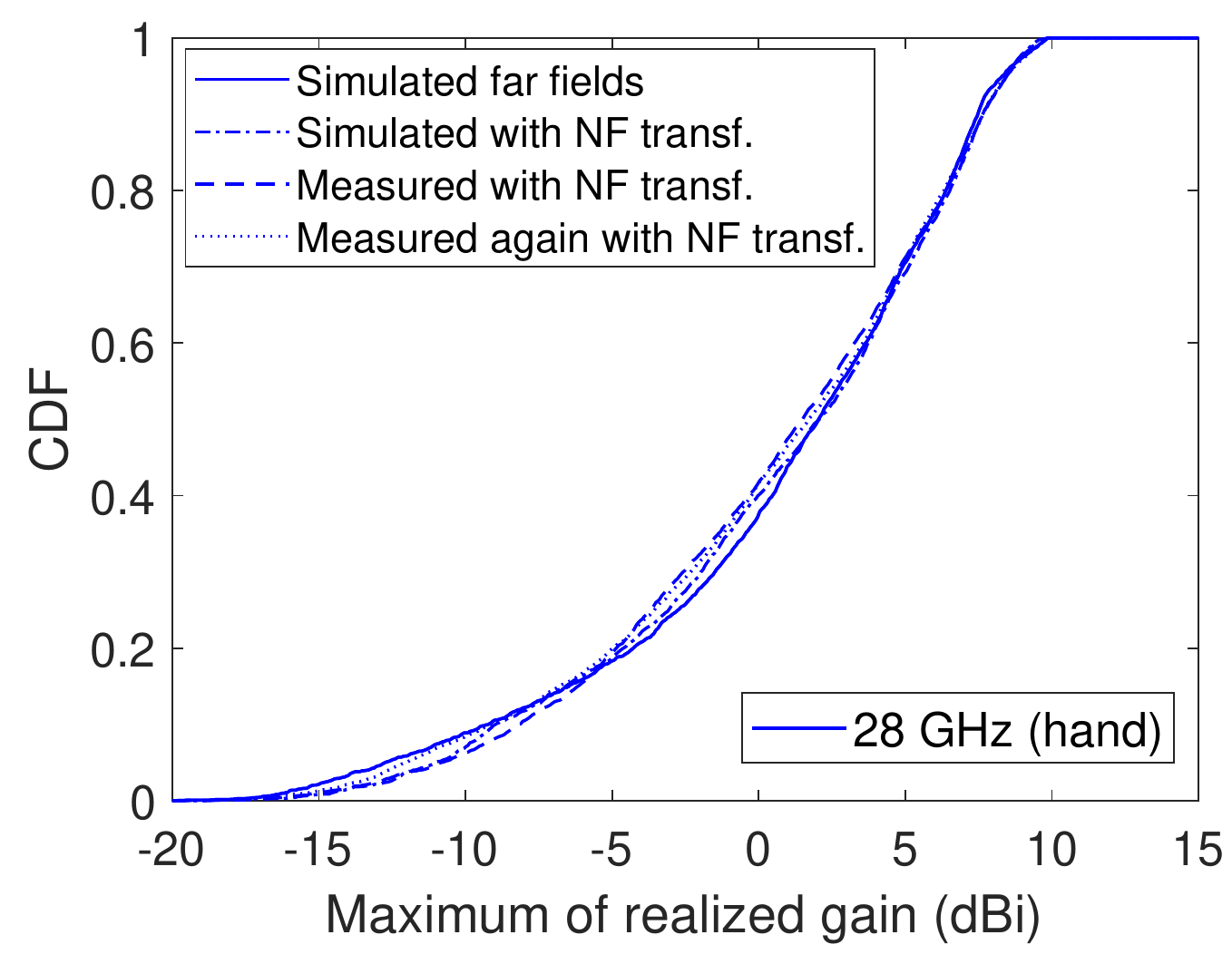}
	\caption{The CDF of the spherical coverage for the 28GHz phone mock-up held by a hand. 'NF transf.' represents near-field to far-field transformation.}
	\label{28handCDF}
\end{figure}
\begin{figure}[htbp] 
    \centering
	\subfigure[]{
	\includegraphics[width=1\linewidth]{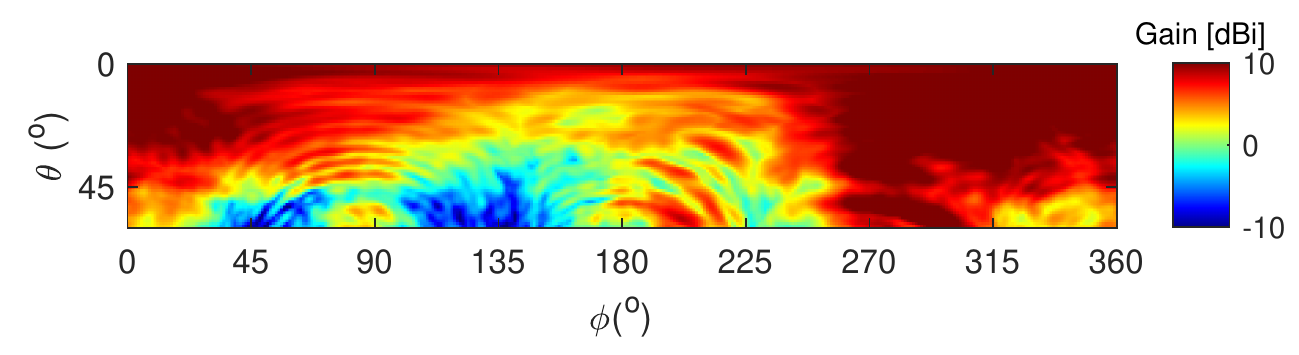}}
	\subfigure[]{
	\includegraphics[width=1\linewidth]{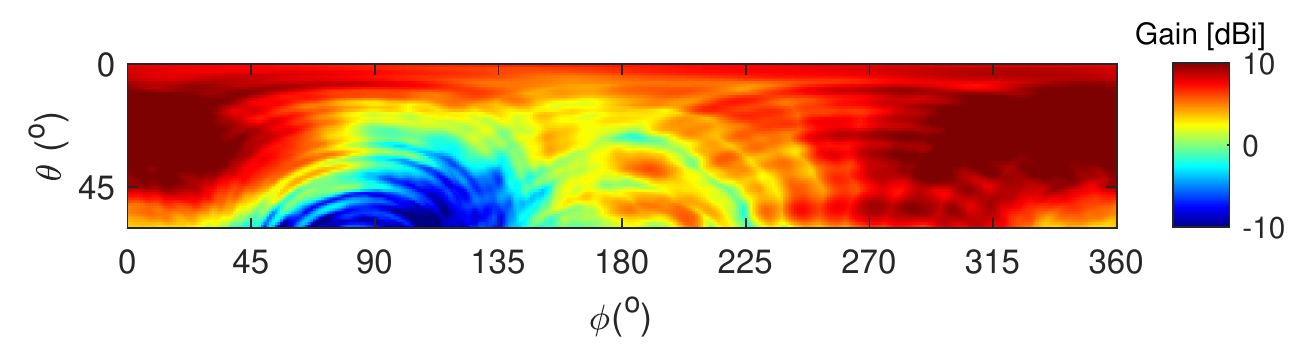}}
	\subfigure[]{
	\includegraphics[width=1\linewidth]{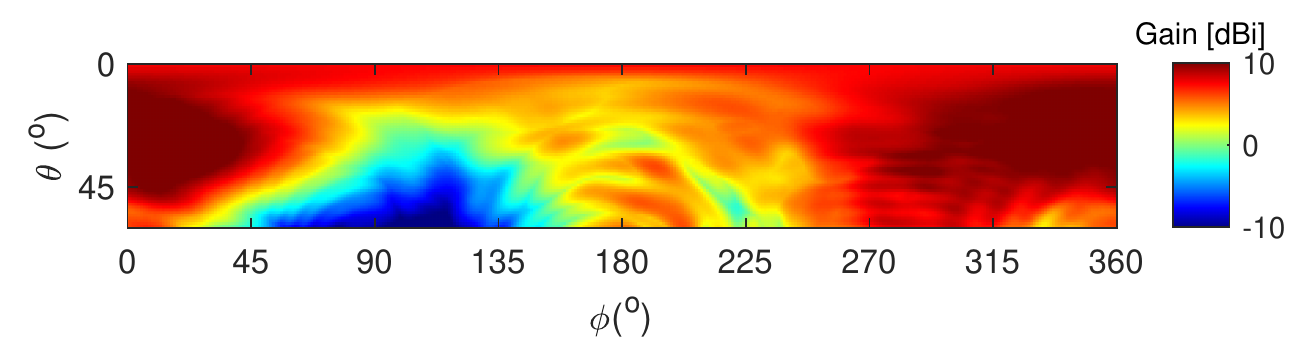}}
	\caption{Spherical coverage of the realized gains of the 39GHz phone mock-up when held by an operator.  (a) Measurement, (b) near-field simulation with transformation and (c) far-field simulation. }
	\label{39handsphericalcoverage}
\end{figure}
\begin{figure}[htbp] 
    \centering
	\includegraphics[width=0.8\linewidth]{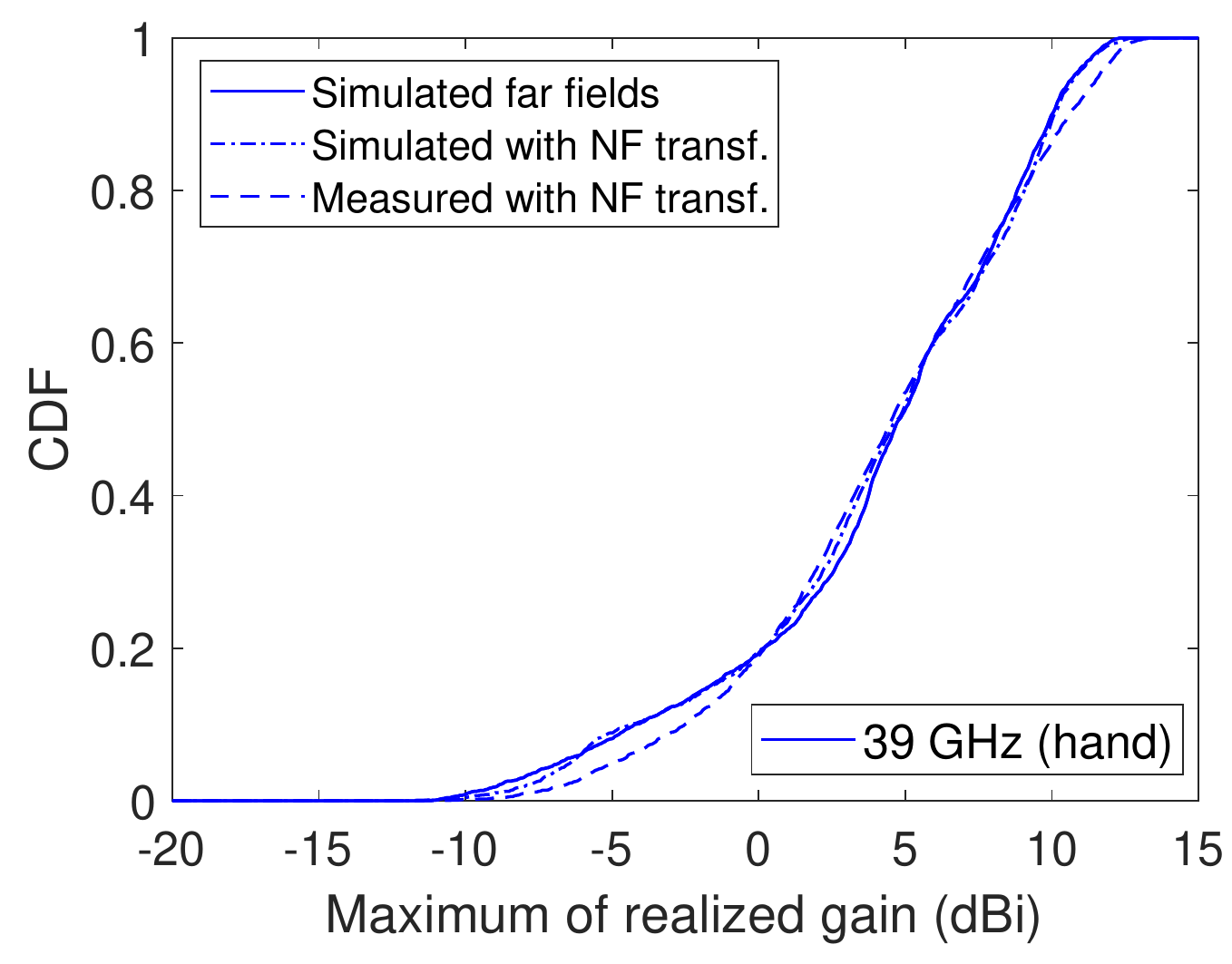}
	\caption{The CDF of spherical coverage for the 39GHz phone mock-up held by a hand. 'NF transf.' represents near-field to far-field transformation.}
	\label{39handCDF}
\end{figure}
\subsection{Measurement Setups for Arrays Held by a User}
The measurement setups for radiation patterns with hand effects differ from those for the free-space case. This is because some parts of a hand are close to antenna elements in the cellphone mock-up, thereby needing some space between them to avoid the detuning effects. Therefore, the distance $d$ should be larger than the free-space measurements. Moreover, hand-tremble effects during the measurements are inevitable for real human bodies that could influence the accuracy of the radiation pattern measurements. The measurement time for each antenna port should be as short as possible to reduce the hand-tremble effects while maintaining the accuracy. To this end, the cellphones are fixed either vertically or horizontally next to the near-field scanner so that only the major polarization of each antenna is measured using the probe. While both $E_x$ and $E_y$ components are measured at the probe in the free-space case, in the antenna-hand interaction measurements, only one of them corresponding to the major polarization is measured. The other polarization component is expected to be much weaker than the major polarization due to the vertical or horizontal fixing of the cellphone mockup. Parameter settings of the near-field scanning are presented in TABLE \ref{setups1}.

This elaborated setup allows us to measure each port within 12 and 15 minutes for 28 GHz and 39 GHz, respectively. The following additional measures are taken to keep the hand posture stable and minimize the impact of varying permittivity of human skin from one part of a hand to another and from one human subject to another:
\begin{enumerate}{}{}
\item{\it Plastic supporters are designed to fix arms;} they are made by a 3D printer using the material with the dielectric constant $<5$ and are put far enough from the antenna array, so that it affects radiations of antennas minimally.
\item{\it Styrofoam supporters are designed to fix wrists;} due to their dielectric constant $\epsilon_{\rm r}\approx 1$, it can be used near the antenna arrays and hardly influences the radiation patterns of antennas.
\item{\it Absorbers are used to reduce unwanted reflections from the metal cable;}
\item{\it If necessary, thin paper tapes are used to fix fingers close to antenna elements and movable parts of hands, like wrists;}
\item{\it Hands are washed using soaps and then dried before the measurements to clean grease on the skin;} and finally,
\item{\it Some pivotal dimensions and locations of the hands and cellphone mockups are marked on them and/or recorded by taking photos;} as each-port measurement requires more than 10 minutes, the human holding the mock-up must have a break before the total 8 ports are covered. The pivotal locations are needed for recovering the postures after each break.
\end{enumerate}
The realized measurement setup is shown in Figs.~\ref{28setups} and \ref{39setups}. After the antenna-hand measurements, 3D models of the hands were obtained. The same postures were recovered by the elaborated setup mentioned above, but a transparent box having the same volume as the cellphone mockups was held to follow the approach in Section~\ref{sec:Modelling}. The generated 3D hand models with a phone chassis are shown in Figs.~\ref{28posture} and \ref{39posture}.
\subsection{Results and Discussions}
Before comparing measured and simulated radiation patterns, implications of measuring only the major polarization of the radiated near electric fields, i.e., either $E_x$ or $E_y$, on the realized accuracy of the pattern estimates are discussed. To this end, a near electric field distribution $\boldsymbol E$ is generated by simulations where the steps $\Delta{x}$ and $\Delta{y}$, the size of the area $S_0$, and the distance $d$ are set according to those of hand measurements in TABLE~\ref{setups1}. By using the near- to far-field transformation introduced in Section~\ref{sec:principle} and applying the loss de-embedding elaborated in Section~\ref{sec:deembedding}, the spherical coverage of $28$~GHz cellphone mock-up yielded a plot in Fig.~\ref{28handsphericalcoverage}. A plot derived from simulated far fields by the CST Studio is also included in Fig.~\ref{fig:28_simFF}, which bypasses our transformation from near- to far-fields. While Fig.~\ref{fig:28_simNF} shows more ripples than Fig.~\ref{fig:28_simFF} because of considering only the major polarization, the gain distributions are quite similar. Plots from the two repeated measurements in Figs.~\ref{fig:28_meas1} and~\ref{fig:28_meas2} also indicate ripples as in Fig.~\ref{fig:28_simNF}. However, the gain distributions of measurements are still like the two plots from simulations. Comparisons of their CDF plots are shown in Fig.~\ref{28handCDF}. There is up to $0.3$~dB difference between the two simulation results, indicating that consideration of only the major polarization does not deteriorate the accuracy of estimating the spherical coverage CDF. Differences between the two repeated measurements and the two simulations are less than $0.6$ and $1$~dB at the median and $0.1$ levels of the CDF. The two measurements show differences of smaller than $0.5$~dB in the CDF, showing repeatability of the measurement setup despite involving non-repeatable human hands.

The spherical coverage of the 39 GHz cellphone mock-up is calculated in Fig. \ref{39handsphericalcoverage}. Like $28$~GHz, there are more ripples on the spherical coverage simulated from the near-field data than that from the far-field data. The plot from the measurement shows similar gain distributions as simulations, although there are more ripples like the plot derived from the near-field simulation. More ripples are in general observed both on $28$ and $39$~GHz plots when derived from the near-field distributions. CDF plots of spherical coverage in Fig.~\ref{39handCDF} show a negligible difference between the two simulations, but more differences between simulations and measurements. At $0.9$ and $0.1$ levels, the differences are about $1$~dB. 
\section{Conclusions}
\label{sec:conclusion}
Impacts of real hands of a cellphone operating person on radiation properties of antenna arrays at two millimeter-wave frequencies, 28 and 39 GHz, are explored. For the first time in the literature, electromagnetic models of antenna arrays including hands are verified against measurements. The cellphone antenna arrays are designed so that they have $2$~GHz impedance bandwidth and $20$~dB isolation between antenna ports. Full-complexity cellphone models including feed lines and cable connectors along with their simplified models with no feed lines and only discrete ports serve different purposes in the antenna evaluation; the former is used for the antenna-hand interaction measurements, while the latter is a reference for de-embedding losses of the antenna measurement setup. Both in the free-space and hand-involved cases, the similarity between simulated and measured spherical coverage characteristics was observed. The measurement setup required special attention when involving human hands because of their non-repeatable nature. For example, far-field measurements were not feasible in our anechoic chamber and hence we resorted to the near-field characterization of the radiated fields. Furthermore, phone mock-ups were fixed at specific postures, i.e., either horizontally or vertically, so that the necessary electric near-field distribution was measured with manageable time duration for a human operator to stay still. The validity of the elaborated near-field measurement setup of antenna arrays with real human hands was confirmed by the mentioned comparisons of the spherical coverage characteristics between simulations and measurements. Their CDF from repeated antenna-hand measurements revealed up to a 0.5 dB difference at the median level. 

\section*{}


\bibliographystyle{IEEEtran}
\balance
\footnotesize
\bibliography{source.bbl}

\end{document}